\documentclass[11pt,a4paper]{article}

\usepackage{jcappub}
\usepackage{dcolumn}
\usepackage{amsfonts}
\usepackage{amsbsy}
\usepackage{hyperref}
\usepackage{rotating}
\usepackage[english]{babel}

\title{No need for dark matter in galaxy clusters within Galileon theory}

\author[a]{Vincenzo Salzano,}
\affiliation[a]{Institute of Physics, University of Szczecin, Wielkopolska 15, 70-451 Szczecin, Poland}
\emailAdd{enzo.salzano@wmf.univ.szczecin.pl}
\author[b]{David F. Mota,}
\affiliation[b]{Institute of Theoretical Astrophysics, University of Oslo, 0315 Oslo, Norway}
\emailAdd{d.f.mota@astro.uio.no}
\author[a,c,d]{Mariusz P. D\c{a}browski,}
\affiliation[c]{National Centre for Nuclear Research, Andrzeja So{\l}tana 7, 05-400 Otwock, Poland}
\affiliation[d]{Copernicus Center for Interdisciplinary Studies, S{\l}awkowska 17, 31-016 Krak{\'o}w, Poland}
\emailAdd{mpdabfz@wmf.univ.szczecin.pl}
\author[e,f]{and Salvatore Capozziello}
\affiliation[e]{Dipartimento di Fisica ``E. Pancini'' , Universita' degli Studi di Napoli ``Federico II'' and  INFN, Sezione di Napoli, Complesso
Universitario di Monte S. Angelo, Via Cinthia, Edificio N, 80126 Napoli, Italy}
\affiliation[f]{Gran Sasso Science Institute (INFN), Viale F. Crispi, 7, I-67100
L'Aquila, Italy}

\emailAdd{capozzie@na.infn.it}

\abstract{Modified gravity theories with a screening mechanism have acquired much interest recently in the quest for a viable alternative to General Relativity on cosmological scales, given their intrinsic property of being able to pass Solar System scale tests and, at the same time, to possibly drive universe acceleration on much larger scales. Here, we explore the possibility that the same screening mechanism, or its breaking at a certain astrophysical scale, might be responsible of those gravitational effects which, in the context of general relativity, are generally attributed to Dark Matter. We consider a recently proposed extension of covariant Galileon models in the so-called ``beyond Horndeski'' scenario, where a breaking of the Vainshtein mechanism is possible and, thus, some peculiar observational signatures should be detectable and make it distinguishable from general relativity. We apply this model to a sample of clusters of galaxies observed under the \textit{CLASH} survey, using both new data from gravitational lensing events and archival data from X-ray intra-cluster hot gas observations. In particular, we use the latter to model the gas density, and then use it as the only ingredient in the matter clusters' budget to calculate the expected lensing convergence map. Results show that, in the context of this extended Galileon, the assumption of having only gas and no Dark Matter at all in the clusters is able to match observations. We also obtain narrow and very interesting bounds on the parameters which characterize this model. In particular, we find that, at least for one of them, the general relativity limit is excluded at $2\sigma$ confidence level, thus making this model clearly statistically different and competitive with respect to general relativity.}


\begin{document}

\maketitle

\section{Introduction}
\label{sec:Introduction}

Despite the astonishing successes which have been accomplished very recently \citep{GW1,GW2}, following the high-level-accuracy cosmological observations collected so far \citep{PlanckCosmo,PlanckMod}, and meanwhile waiting for newest and greatly improved projects to be launched and fully operative (SKA\footnote{\url{https://www.skatelescope.org/}} and \textit{Euclid}\footnote{\url{http://sci.esa.int/euclid/}} \citep{Laureijs09,Laureijs11,Refregier10,Amendola13} among others), the full validity of general relativity (GR) over all the possible astrophysical and cosmological ranges is still under debate (see \citep{BeyondLCDM} and reference therein). This is mainly due to the enigmatic nature of two of its main pillars which are nowadays believed to rule the dynamics of the entire universe and of everything is inside it: Dark Matter (DM) and Dark Energy (DE). It is indisputable that no candidate for DM has yet been detected; some break in the Standard Model of Physics might be possible, as stated in \citep{LHC1,LHC2,LHC3}, but the results have not been statistically confirmed yet, and they might be only loosely related to the DM problem. Also, we are basically unaware about the nature of DE \citep{DE1,DE2} and willing for more decisive statistical evidences from observations \citep{DE3}.

What could be good or bad news is that we have plenty of alternatives to GR, at least for what concerns the interpretation and the solution of DE-related problems \citep{BeyondLCDM,Clifton12,clifton}. And all can be bounded, and many times disproved, by requiring GR as a limit at such scales where we know it works well as, for example, Solar System scales \citep{Will14}. The family-theories that generally pass this test are based on some sort of \textit{screening mechanism}: the new introduced degrees of freedom (whatever they are, in the form of a new scalar field, or as geometrical effects) are generally suppressed at small scales, so that any deviation from GR cannot be detected \citep{Hamilton15,shaw,gano}, but stay active at larger cosmological scales, where they can mimic DE, with no ad-hoc assumptions and in a theoretical apparatus possibly much more general than GR. Even in this case, anyway, we have a generous selection of mechanisms of various origin which might produce such screenings (see \citep{Joyce15,bour,Berti15} for a general review).

Here, we will focus on a well-defined scenario: a Galileon-type alternative model, which spontaneously breaks the underlying Vainshtein screening mechanism. Galileons are a particular class of scalar field invariant under galilean shift symmetry and which, despite of leading to higher-derivative field theories, still have second order equations of motions \citep{Nicolis09,Deffayet09,Deffayet11,Deffayet13}. Generally, they are able to pass Solar System scale test by means of a the so-called Vainshtein mechanism \citep{Vainshtein72} due to the particular way the kinetic contributions of the Galileon fields are defined, with first or second order derivatives becoming important at a certain scale.

Such a theory has been widely studied; actually, we will focus on a particular version of it, recently proposed in \citep{KoyamaSakstein2015}. This model is very interesting because it naturally contains a breaking of the Vainshtein mechanism at some (to be defined) scale, which basically implies that there should be some breaking of GR at some point, and this should be detectable \cite{mauro}. One possible way to test such a hypothesis has been proposed in \citep{SaksteinPRL15,SaksteinPRD15} in a stellar physics context. A generalized version of it has been discussed in \citep{Sakstein2016} and tested with clusters of galaxies, using both gravitational lensing and X-ray observations.

Here, we will try to face a very different approach: such a model has been considered in all the above cases only as a general alternative to GR, in particular, as a good candidate for DE, but no suggestion about a possible role played in substituting DM has been given. In general, most of these alternative theories are suggested in order to explain the dynamics of the universe on cosmological scales without requiring the presence of an exotic fluid like DE. In the specific case of Galileon models, they have been studied in details as cosmological background and, as such, influencing the formation of gravitational structures. But no direct connection with a possible role in mimicking DM has been defined. Actually, even theoretically, such possibility was in some way prohibited by the same screening mechanism. But with the new model developed in \citep{KoyamaSakstein2015,Sakstein2016} we have a natural breaking of such mechanism so that one could naturally ask: \textit{what if the mechanism were broken at some astrophysical scale(s) and, thus, the Galileon model might play the role of DM in some way?}

In order to give an answer to this question, we will approach the problem in the following way. First, we have to identify the kind of observations which might result to be helpful for our purposes. We found it in the convergence map that can be derived from the analysis of strong and weak lensing events from clusters of galaxies. Actually, clusters of galaxies are the best gravitational objects in order to combine both phenomena: they constitute massive and large enough foreground lenses able to produce detectable distortions of background sources both near the center (where the depth of the gravitational potential creates strong lensing events) and in the outer regions (where the cluster mass produces the statistical distortions of background galaxies known as weak lensing). Finally, the convergence map will result to be, basically, the two-dimensional projected mass distribution of the clusters that can be reconstructed by taking into account all these lensing events.

Clusters of galaxies are also characterised by another dynamical property: the large amount of intra-cluster gas (the largest matter component, if we do not consider dark matter) is heated up by the deep gravitational potential and emits in the X-ray band. X-ray observations of hot gas in clusters of galaxies provide us with another tool to have information about the internal mass distribution of a cluster. As it is well known, there might be internal astrophysical phenomena which can locally perturb the gas, or even larger-scale events like merging from smaller sub-structures, which can heat it up in addition to the pure gravitational attraction, leading to not properly correct mass estimations. Mass reconstruction from lensing, on the other hand, is much more accurate because it is insensitive to local dynamics, and can reproduce the true mass distribution due to pure gravitational attraction quite well.

Actually, we will not be interested directly on these aspects in this work, but more on the fact that, through X-ray observations, the gas density distribution of the cluster can be measured with very high confidence. Such gas density will be then used as input to calculate the convergence map from lensing in the context of the Galileon model we have discussed above. The important point to be stressed is that the gas will be the only contribution to the cluster mass we will consider; no additional DM will be assumed here.

The paper is organized as follows: in section~(\ref{sec:Model}) we describe the theoretical apparatus at the base of our analysis; in section~(\ref{sec:Data}) we describe what kind of data we have considered and how we implement them; in section~(\ref{sec:Results}) we discuss the results obtained and their implications; finally, in section~(\ref{sec:Conclusions}), we summarize all our analysis.

\section{The Extended Horndeski Model}
\label{sec:Model}

We will not give all the details of the model introduced in \citep{KoyamaSakstein2015,Sakstein2016}, but we will only describe the main elements we need for our analysis; the interested reader may find more insights in the above references. While the original Galileon models are an example of Horndeski theory \citep{Horndeski1974}, i.e. the most general scalar-tensor theory that gives rise to manifestly second-order field equations for both the scalar and the metric, recently it has been shown that it is possible to extend Horndeski theories in some ``healthy'' ways, which still have second order equation of motion (manifestly or not) \citep{Zumalacarregui14,Gleyzes14,Gleyzes15A,Gleyzes15B}. Models in \citep{KoyamaSakstein2015,Sakstein2016} belong to this so-called ``beyond Horndeski'' theories and are defined in \citep{KoyamaSakstein2015} through the Lagrangian
\begin{equation}
\frac{\mathcal{L}}{\sqrt{-g}} = M^2_{Pl} \left[ \frac{R}{2} - \frac{1}{2} \partial_{\mu}\phi \partial^{\mu}\phi + \frac{\mathcal{L}_{4}}{\Lambda^4} \right]\; ,
\end{equation}
where $g$ is the determinant of the metric; $R$ is the Ricci scalar; $\mathcal{L}_4$ is defined as
\begin{equation}
\mathcal{L}_{4} \equiv -X \left[ \left(\square \phi \right)^{2} - \phi_{\mu\nu} \phi^{\mu\nu} \right]
- \left( \phi^{\mu} \phi^{\nu} \phi_{\mu\nu} \square \phi - \phi^{\mu} \phi_{\mu\nu} \phi_{\rho} \phi^{\rho\nu}\right)\; ,
\end{equation}
where $\phi$ is the Galileon field; $\phi_{\mu_{1}\ldots\mu_{n}} \equiv \nabla_{\mu_{1}}\ldots\nabla_{\mu_{n}}\phi$ and $X \equiv -1/2 \partial_{\mu}\phi \partial^{\mu}\phi$ is the standard kinetic term; and the reduced Planck mass is defined as $M_{Pl} = (8\pi G)^{-1}$, where $G$ is the bare gravitational constant and differs from the usually measured one, $G_{N}$. Assuming a metric signature $(-, +, +, +)$, and the Newtonian Gauge, the perturbed Friedmann-Lema\^{i}tre-Robertson-Walker (FLRW) metric can be written as
\begin{equation}\label{eq:metric_perturb}
\mathrm{d} s^2 =-\left[1+2\frac{\Phi(r,t)}{c^2}\right] c^2 dt^2+a^2(t)\left[1-2\frac{\Psi(r,t)}{c^2}\right]\delta_{ij} \mathrm{d}x^i \mathrm{d}x^j\; ,
\end{equation}
where $c$ is the speed of light; $a$ is the cosmological scale factor; and $\Phi$ and $\Psi$ are the gravitational and the metric potentials for an extended object. Thus, the model can be fully described by the following equations \citep{Sakstein2016}:
\begin{equation}\label{eq:potential_phi}
\frac{\mathrm{d} \Phi(r)}{\mathrm{d} r} = \frac{G_{N} M(r)}{r^2} + \frac{\Upsilon_{1}}{4} G_{N} M''(r) \; ,
\end{equation}
\begin{equation}\label{eq:potential_psi}
\frac{\mathrm{d} \Psi(r)}{\mathrm{d} r} = \frac{G_{N} M(r)}{r^2} - \frac{5\Upsilon_{2}}{4} \frac{G_{N} M'(r)}{r} \, ,
\end{equation}
where $M(r)$ is the mass enclosed in a radius $r$, primes denote derivative with respect to radius $r$, and $\Upsilon_{1}$ and $\Upsilon_{2}$ are two constants which completely characterise this model, and quantify the breaking of the Vainshtein mechanism and, thus, its departure from GR. In particular, one recognises that GR is restored in the limit of $\Upsilon_{1} \rightarrow 0$ and $\Upsilon_{2} \rightarrow 0$.

\section{Data from galaxy clusters}
\label{sec:Data}


The sample of clusters we are going to analyse has been observed by the survey program Cluster Lensing and Supernova survey with Hubble, \textit{CLASH} \citep{Postman12}, a multi-cycle treasury program which, using $524$ Hubble Space Telescope (HST) orbits, has observed (among others) $20$ clusters, selected following the criterium of an approximately unperturbed and relatively symmetric X-ray morphology. For these clusters, we have both X-ray data (see \citep{Donahue14} for more insights), from which we can extract the gas density profiles, and a lensing-based analysis, described in \citep{Merten15}.

\begin{figure*}[htbp]
\centering
\includegraphics[width=\textwidth]{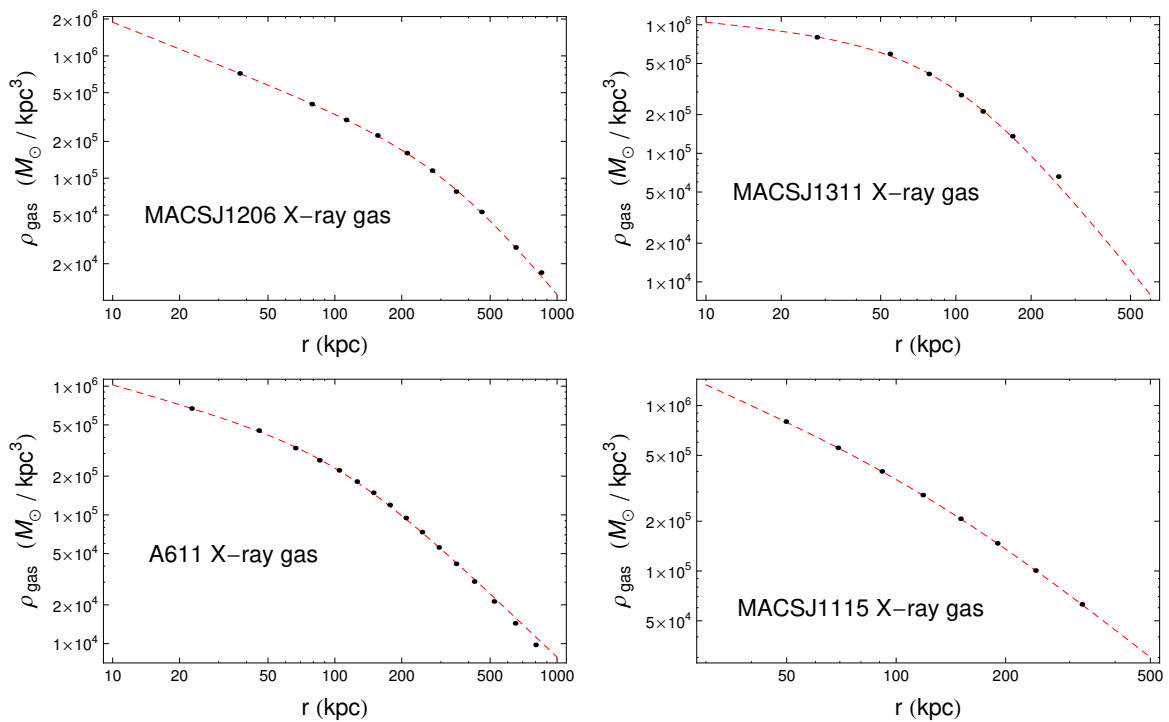}
\caption{Comparison of X-ray derived gas density (black points) with best fit relation Eq.~\ref{eq:gas_density} obtained using the same data.)}\label{fig:Xray_gas}
\end{figure*}

In \citep{Donahue14}, the authors analyse the mass profiles of the clusters using archival data from \textit{Chandra}, which are reprocessed, re-calibrated and analysed using the procedure outlined in \citep{Donahue14}. As we have said in the Introduction, we are interested only in the gas density profiles as can be derived from X-ray surveys: following \citep{Donahue14} we fit such data with a $\beta$-model \citep{Cavaliere78} truncated at low-radius by a power-law,
\begin{equation}\label{eq:gas_density}
\rho_{gas} = \rho_{e,0} \left( \frac{r}{r_{0}} \right)^{-\alpha}  \left[1 + \left(\frac{r}{r_{e,0}}\right)^{2} \right]^{-\frac{3\beta_{0}}{2}} \; ,
\end{equation}
where the truncation is needed in order to match the peculiar features of the inner cores of clusters, and where $\rho_{e,0}$ is the density normalization, $r_{e,0}$ is the typical scale radius for the $\beta$-model, and $r_{0}$ is the scale related to the power-law truncation term added to the same $\beta$-model. This equation will be used as input in the theoretical Galileon framework to calculate the convergence map from lensing analysis: we first use the density gas data available from X-ray surveys and fit them with Eq.~\ref{eq:gas_density}; then, the obtained best fit function is used in the lensing analysis as matter component to calculate gravitational potentials Eqs.~\ref{eq:potential_phi}~-~\ref{eq:potential_psi}. The impact on the final results of fixing the gas parameters at this stage is absolutely negligible, as the X-ray gas density data are perfectly matched by such a model. Possibly, Eq.~\ref{eq:gas_density} might over-estimate gas density at larger distances from the center, but in our sample the truncated $\beta$-model still works very well in this range. As a proof of that, in Fig.~\ref{fig:Xray_gas} we plot some clusters from our sample, comparing observed gas density (black points) with the obtained best fit Eq.~\ref{eq:gas_density}.

The most general expression for the convergence, namely, not dependent on the chosen gravity theory, is
\begin{equation}\label{eq:kappa_general_mod}
\kappa(R) =  \frac{1}{c^{2}} \frac{D_{l}D_{ls}}{D_{s}} \int^{+\infty}_{-\infty} \nabla_{r} \left( \frac{\Phi(R,z) + \Psi(R,z)}{2}\right)\, dz \, ,
\end{equation}
where $c$ is the speed of light; $R$ is the two-dimensional projected radius; $z$ is the line of sight direction; $\Phi$ and $\Psi$ are the total gravitational and metric potentials; and $D_{s}$, $D_{l}$ and $D_{ls}$ are, respectively, the source-observer, the lens-observer and the lens-source angular diameter distances. We have to point out here that the angular diameter distances are calculated as
\begin{equation}
D_{A}(z) = \frac{1}{1+z}\int^{z}_{0} \frac{c \; \mathrm{d}z'}{H(z',\boldsymbol{\theta})} \; ,
\end{equation}
where $z$ is the redshift of the cluster and $H(z,\boldsymbol{\theta})$ is the Hubble function, which depends on the background theoretical model through the vector of parameters $\boldsymbol{\theta}$. We don't have a cosmological analysis for the model we are considering; that is why we use a standard fiducial cosmological model (only to calculate these distances) as given in \citep{Merten15}, i.e., a $\Lambda$CDM model with the dimensionless matter density parameter today $\Omega_{m} = 0.27$, the dimensionless DE parameter today $\Omega_{DE} = 0.73$ (assuming null spatial curvature), and the Hubble constant $H_{0} = 70$ km s$^{-1}$ Mpc$^{-1}$. One could ask whether such a choice could bias or have some influence on the final result. In \citep{Barreira14C}, where a general Galileon model is analyzed, it is found that the rate expansion $H(z,\boldsymbol{\theta})$ does not really depend on Galileon parameters, but only on $\Omega_{m}$ which, furthermore, stays basically unchanged when considering $\Lambda$CDM (in GR) or a Galileon theory, so that its value is fundamentally non influential for our purposes. Moreover, in \citep{Barreira14C} it is also shown that the Galileon Hubble function $H(z)$ exhibits a difference $\lesssim5\%$ with respect to a $\Lambda$CDM at $z\lesssim 1$, which is the redshift range covered by our data. Such a difference is smaller than present observational errors on most of the cosmological data we have so far, thus, statistically speaking, these two (very) different approaches are basically indistinguishable and we cannot exclude that better data would not lead to even a smaller difference between them.

We will also analyze the ``classical'' GR case, where a DM profile is assumed; we will consider the standard Navarro-Frenk-White (NFW) profile \citep{NFW}, given by
\begin{equation}\label{eq:NFW}
\rho_{NFW} = \frac{\rho_s}{(r/r_s)(1+r/r_s)^2}\; ,
\end{equation}
where the only free parameters are a density $(\rho_{s})$ and a scale $(r_{s})$, and we will compare the results with the Galileon case.

In order to test the model, we have to define the $\chi^2$ function, which for the lensing data is
\begin{equation}
\chi^2_{lens} = \boldsymbol{(\kappa^{theo}(\theta)-\kappa^{obs})}\cdot\mathbf{C}^{-1}\cdot\boldsymbol{(\kappa^{theo}(\theta)-\kappa^{obs})} \; ,
\end{equation}
where $\boldsymbol{\kappa^{obs}}$ is the vector of the observationally measured convergence, $\boldsymbol{\kappa^{theo}(\theta)}$ is the theoretical convergence obtained from the right hand side of Eq.~(\ref{eq:kappa_general_mod}) and $\mathbf{C}$ is the corresponding observational covariance matrix. The vector of theoretical parameters will be: $\{\rho_{s},r_{s}\}$ when considering the GR+NFW case, and $\{\Upsilon_{1},\Upsilon_{2}\}$ when we will consider the Galileon+gas scenario. Note that the two models have the same number of parameters.

Finally, in order to establish a well-based statistical comparison, we will calculate the Bayesian Evidence\footnote{There are many other tools which might be used, but each of them is plagued by some problems and that makes the choice of the right one controversial. A list of the most used in cosmology can be
found in \citep{InfoCriteria}.} $\mathcal{E}$ for both the models using the algorithm described in \citep{Mukherjee06}. Then, we will calculate the Bayesian Factor in order to assess a hierarchy of statistically favoured scenarios. This is given by the ratio between the evidences obtained from the two models, i.e., $\mathcal{B}^{Gal.}_{GR} = \mathcal{E}_{Gal.}/ \mathcal{E}_{GR}$. If $\mathcal{B}^{Gal.}_{GR}>1$ than the Galileon model is favoured with respect to GR, and the other way around. Actually, to quantify how much a model is favoured with respect to another is not an easy task; we will follow the Jeffreys' Scale \citep{Jeffreys98} which states that: if $\ln \mathcal{B}^{i}_{j} < 1$, the evidence in favor of model $M_i$ is not significant; if $1 < \ln \mathcal{B}^{i}_{j} < 2.5$, the evidence is substantial; if $2.5 < \ln \mathcal{B}^{i}_{j} < 5$, is strong; if $\mathcal{B}^{i}_{j} > 5$, is decisive. The same scale but with negative values, of course, can be easily interpreted as evidence against model $M_i$.

\section{Results}
\label{sec:Results}

\begin{figure*}[htbp]
\centering
\includegraphics[width=7.6cm]{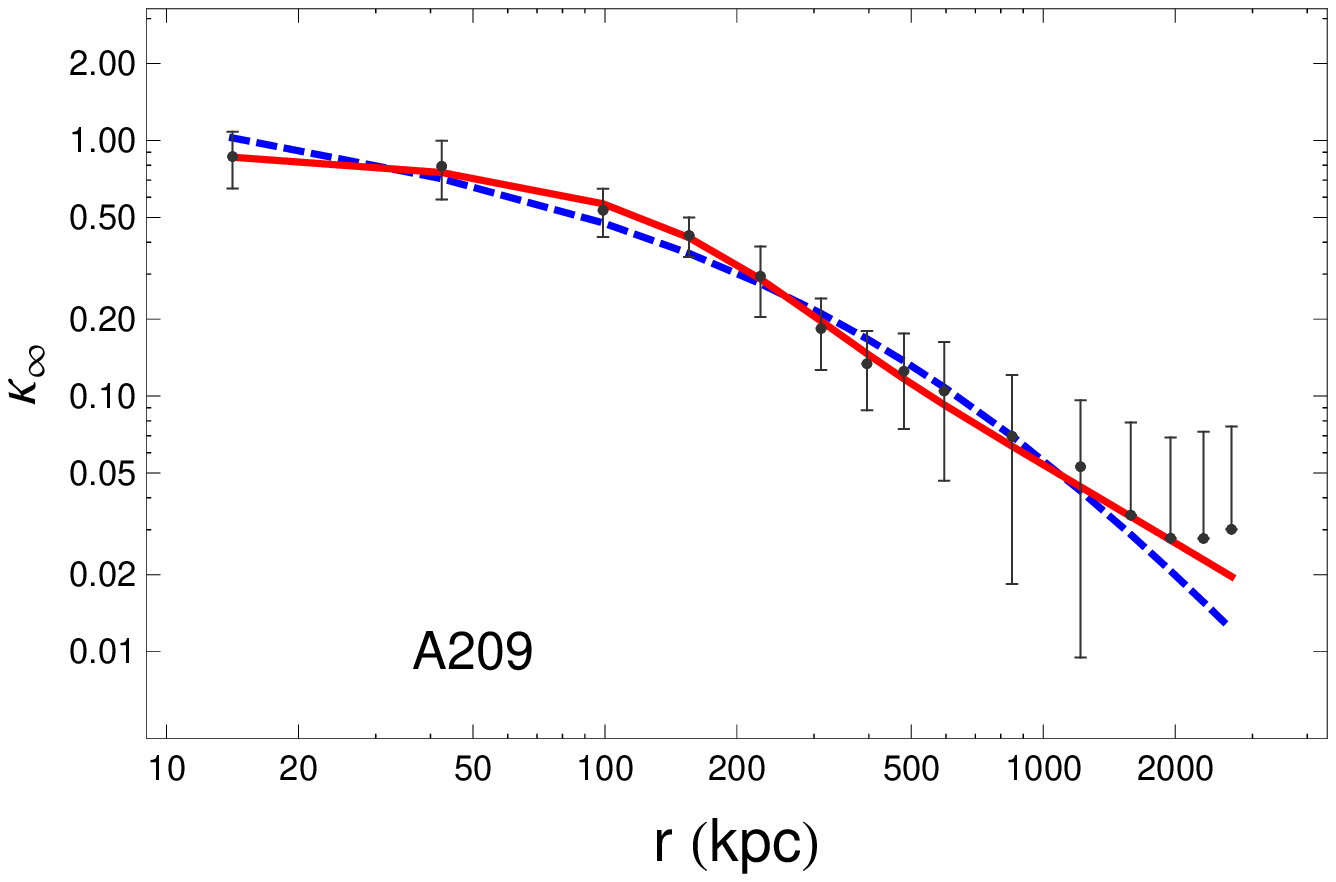}~
\includegraphics[width=7.6cm]{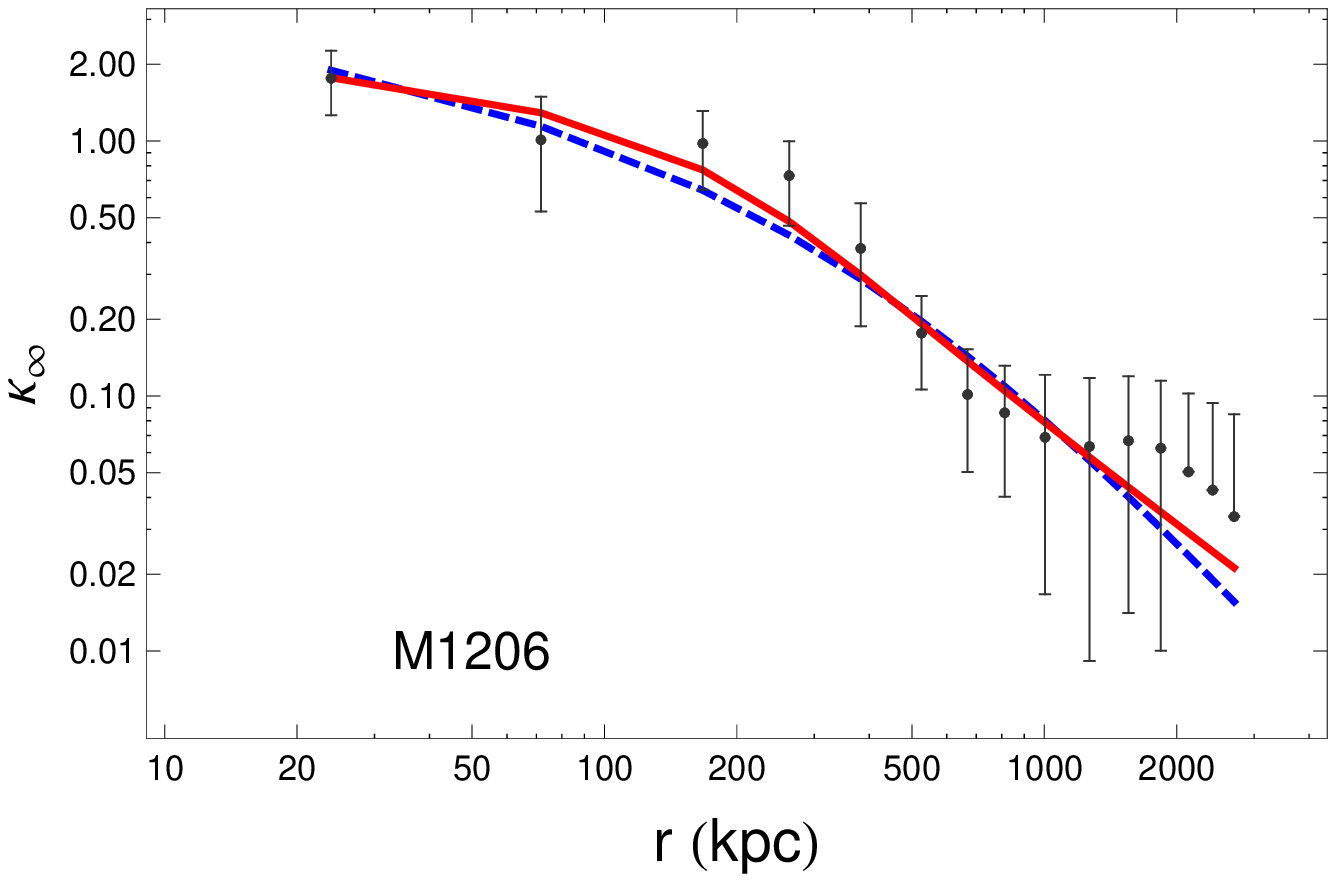}\\
~~~\\
\includegraphics[width=7.6cm]{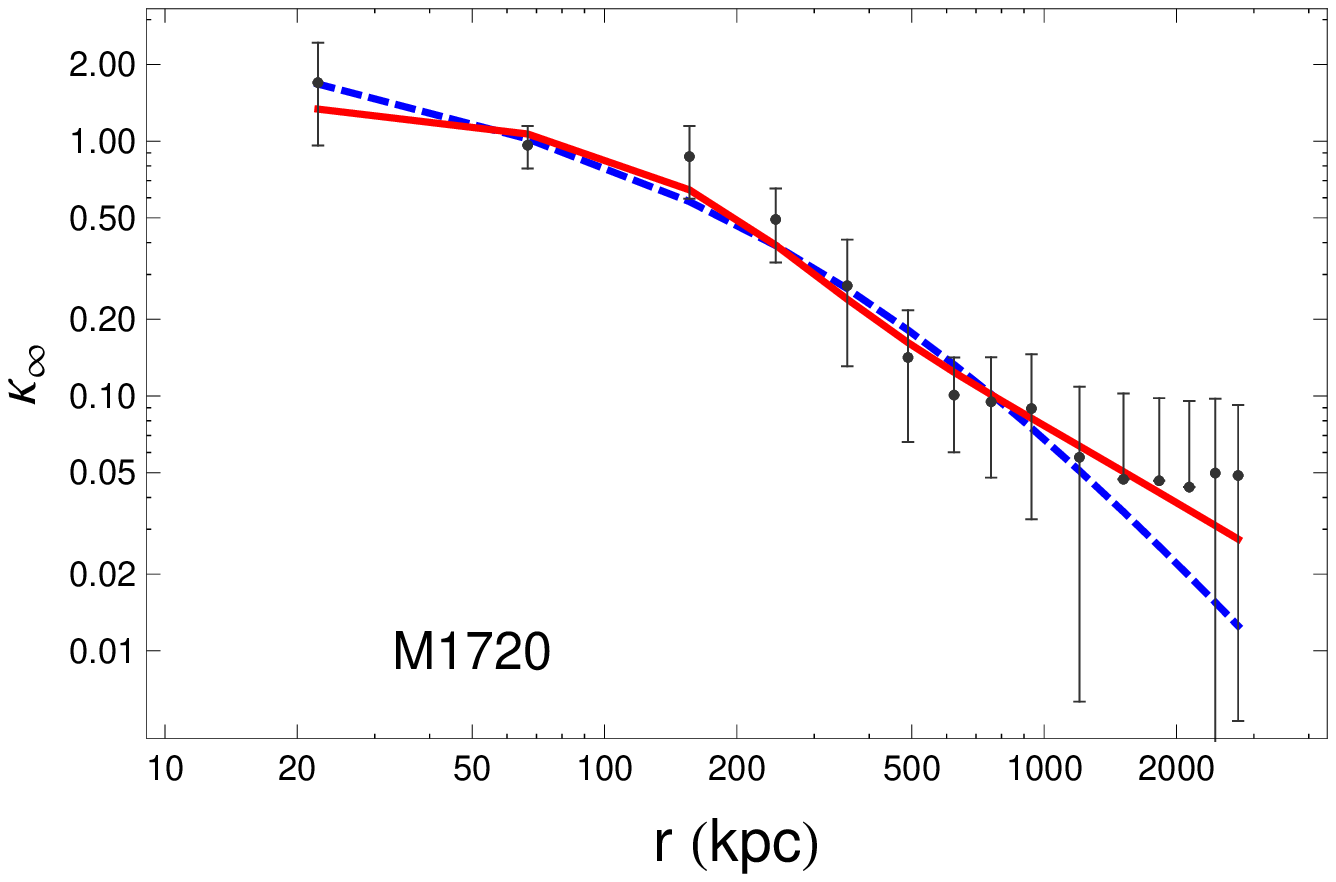}~
\includegraphics[width=7.6cm]{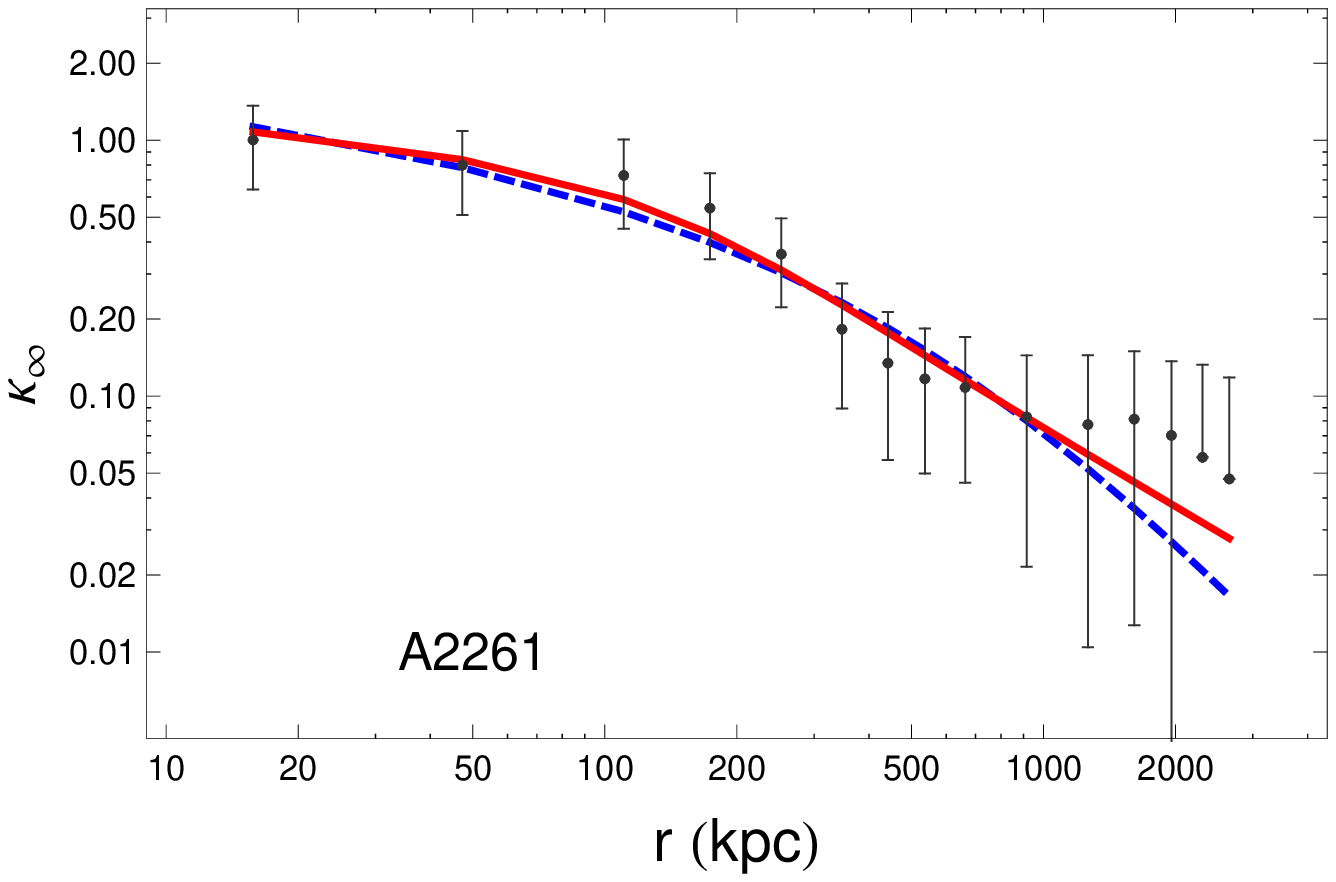}\\
~~~\\
\includegraphics[width=7.6cm]{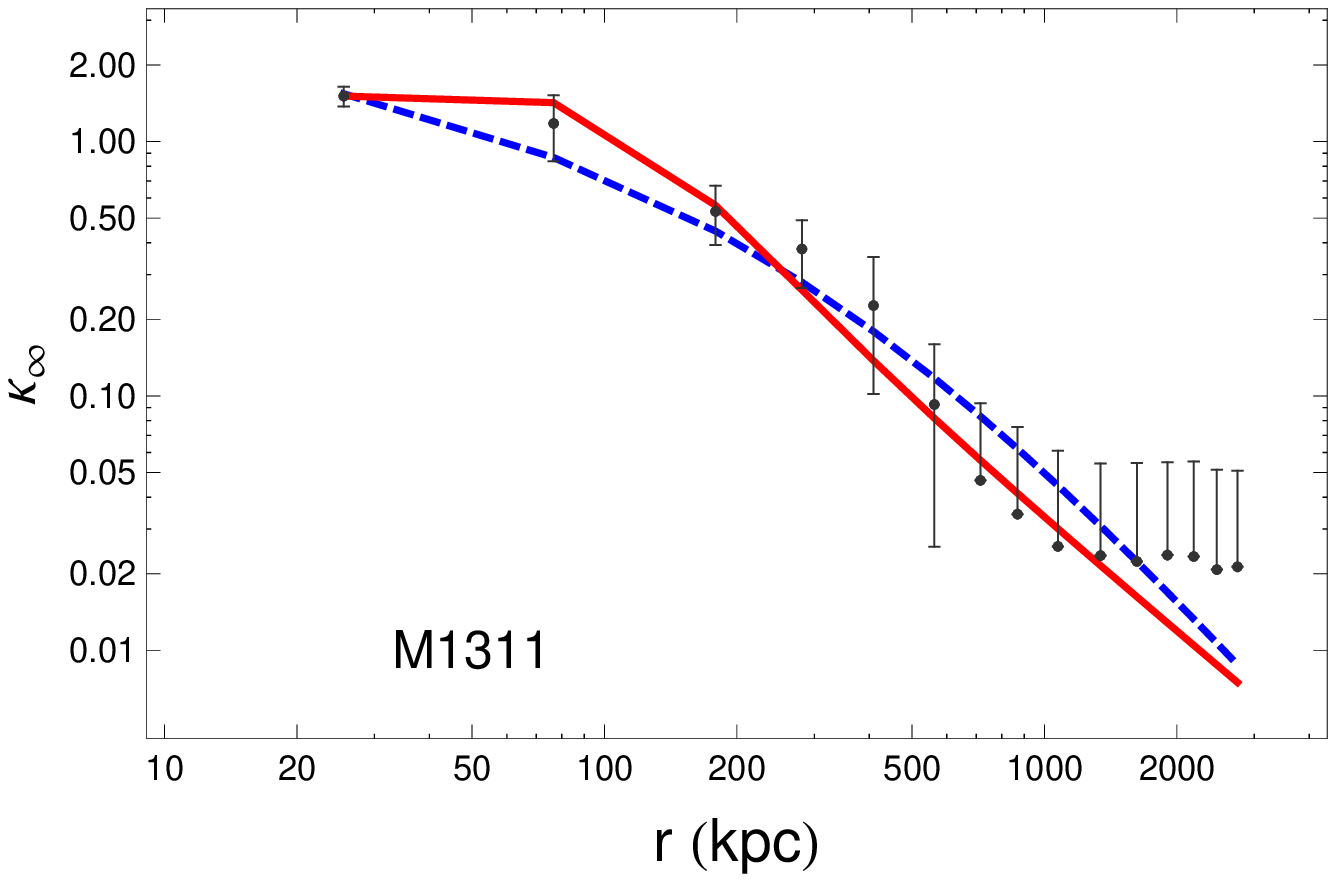}~
\includegraphics[width=7.6cm]{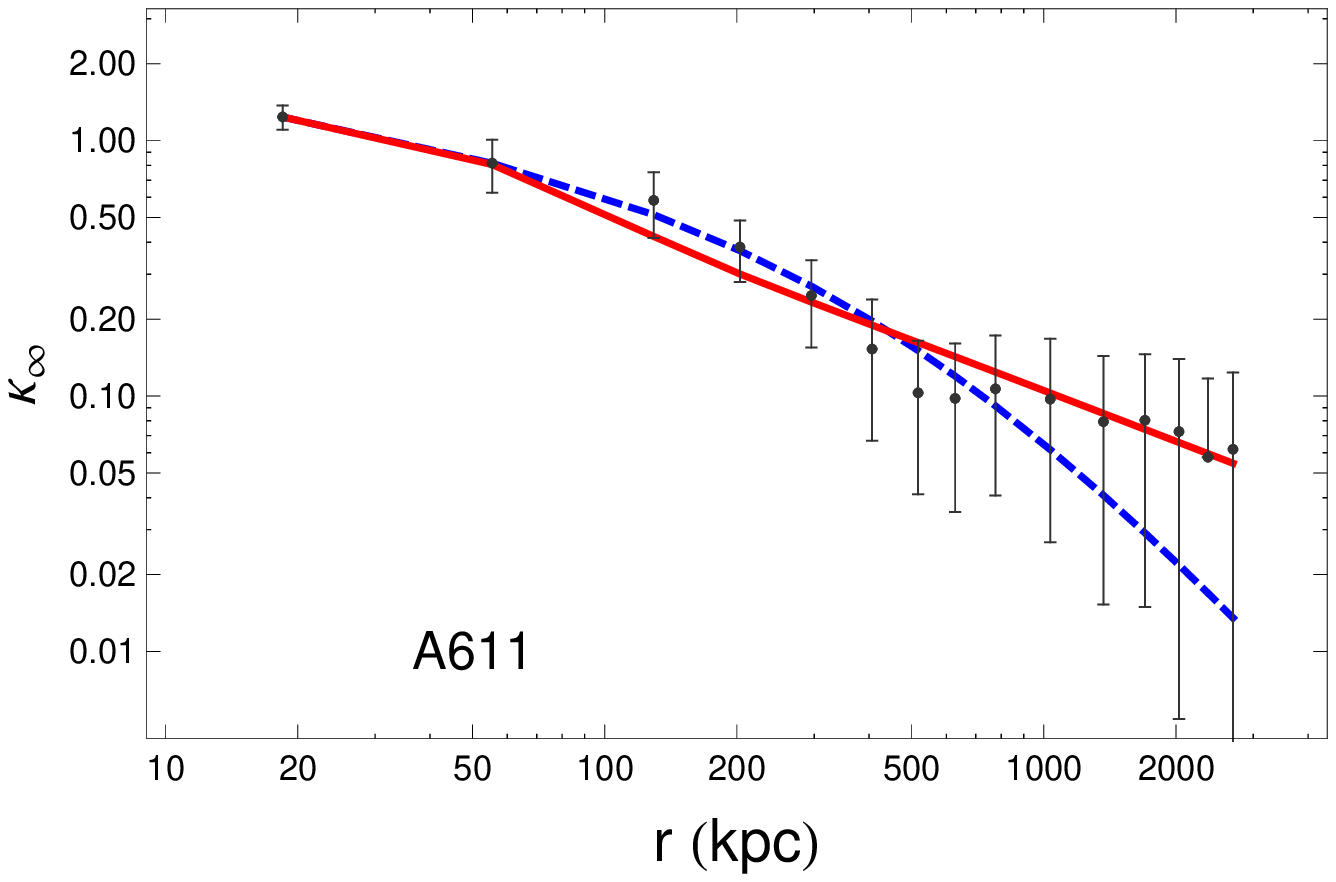}\\
~~~\\
\includegraphics[width=7.6cm]{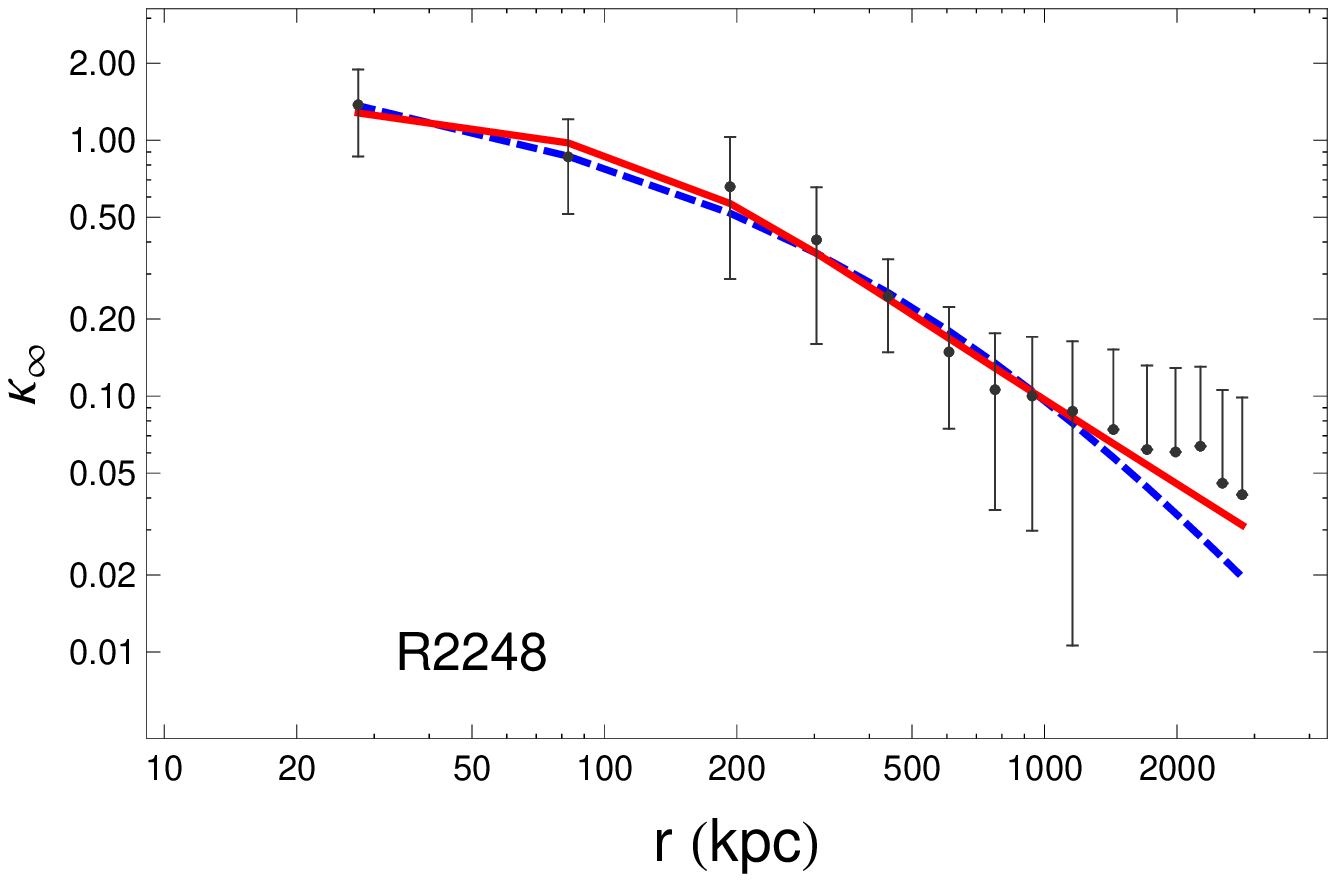}~
\includegraphics[width=7.6cm]{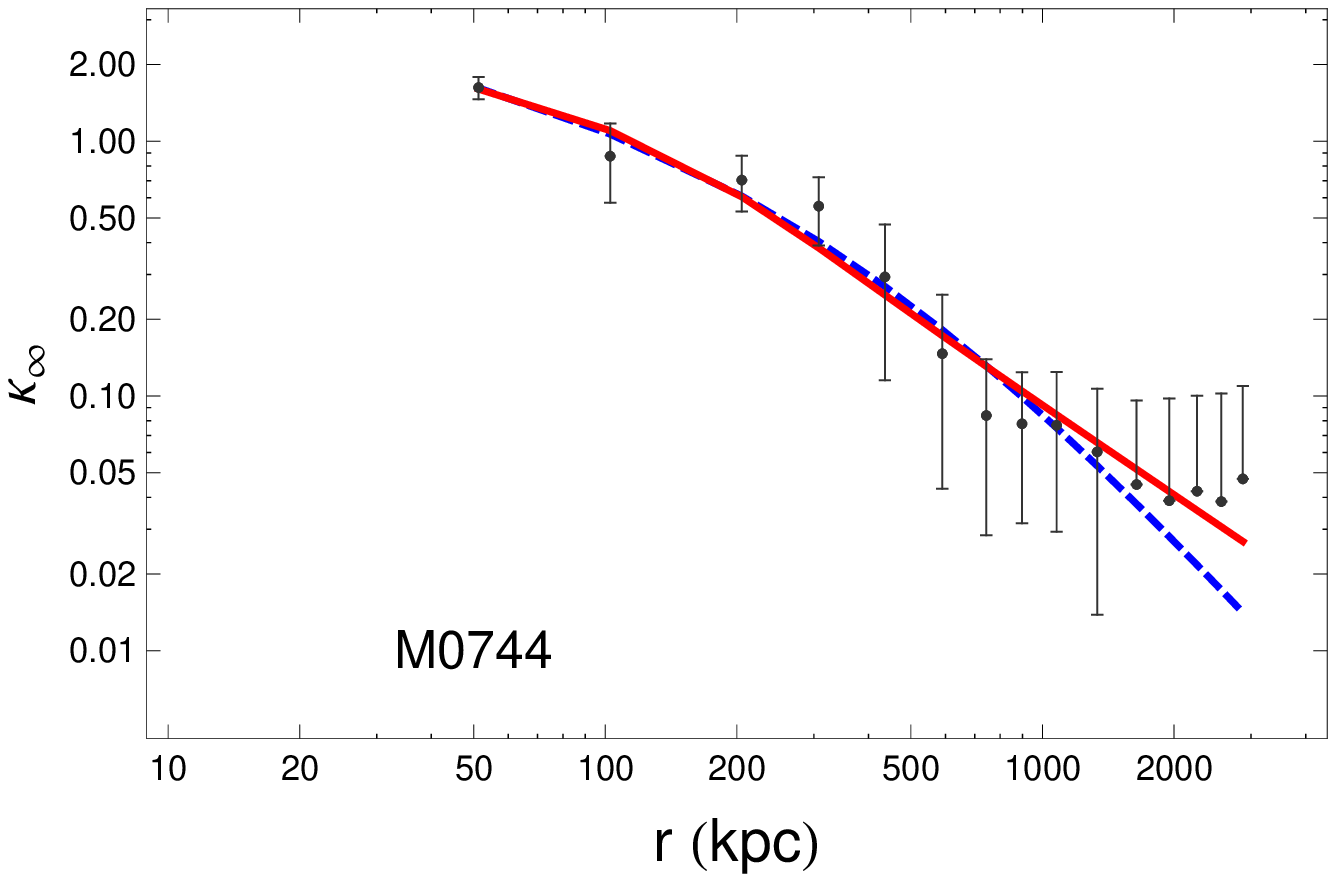}
\caption{Convergence map from gravitational lensing reconstruction. Color code: grey points - observational data; dashed blue - GR + NFW; solid red - Galileon + gas.}\label{fig:fit_1}
\end{figure*}

\begin{figure*}[htbp]
\centering
\includegraphics[width=7.6cm]{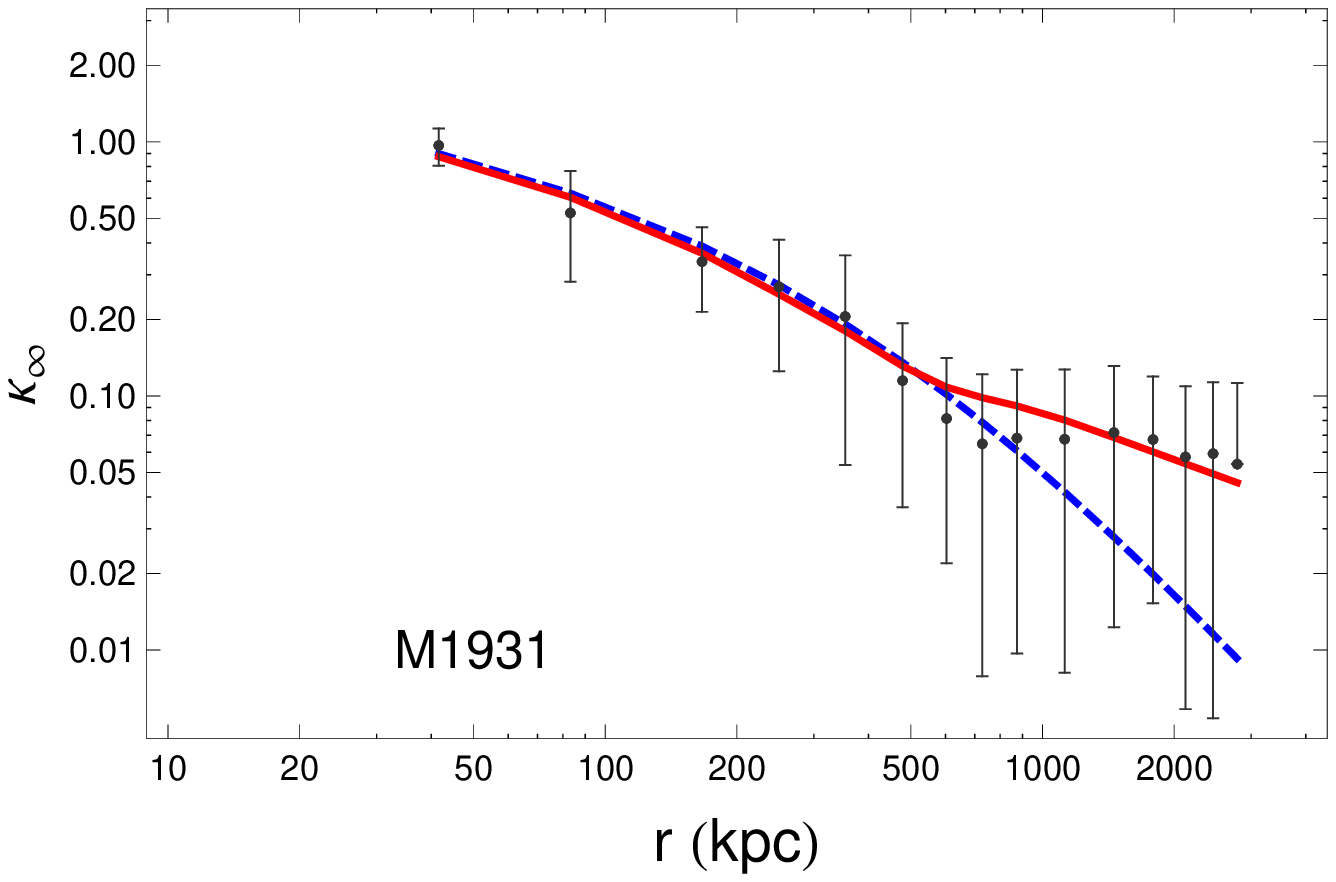}~
\includegraphics[width=7.6cm]{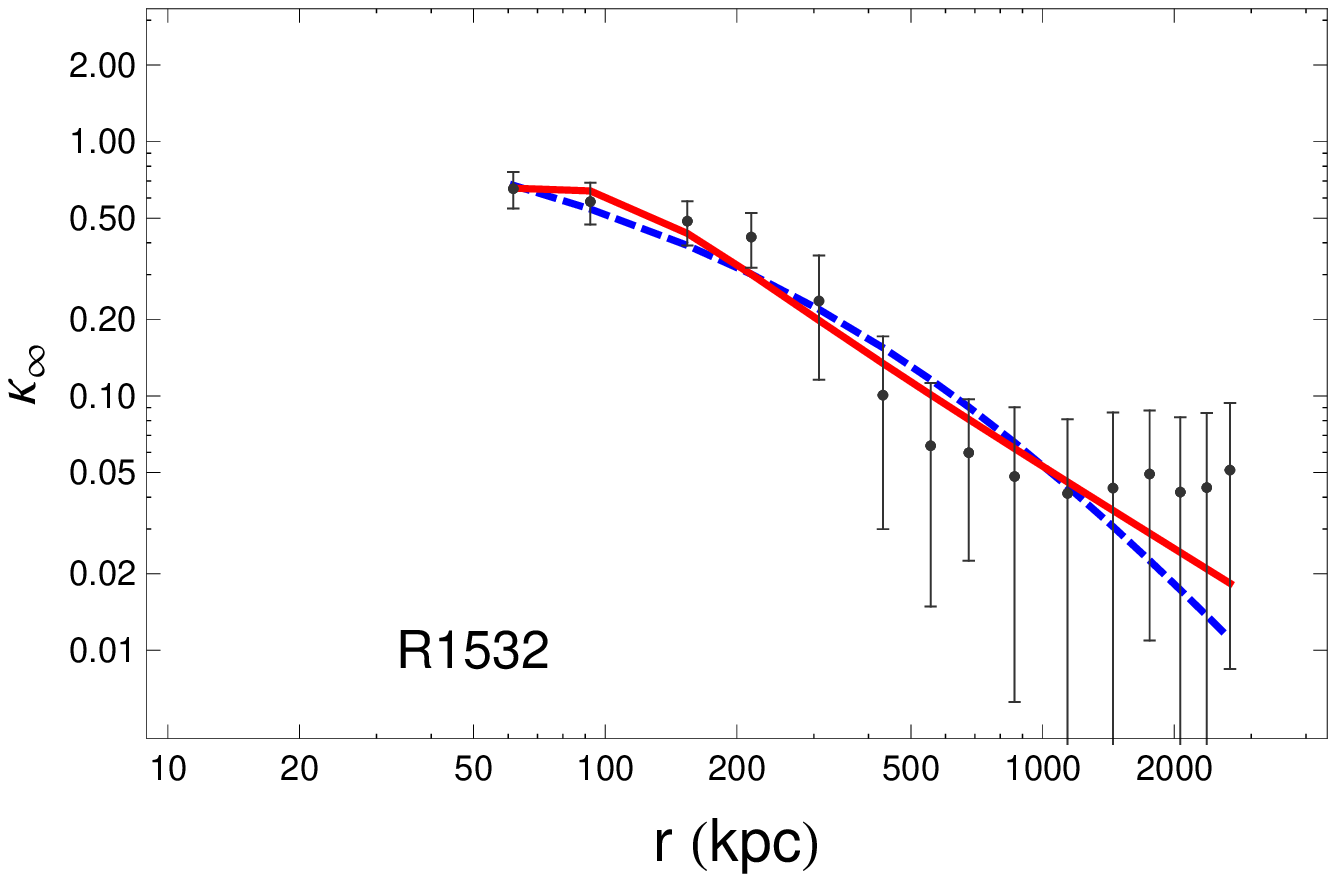}\\
~~~\\
\includegraphics[width=7.6cm]{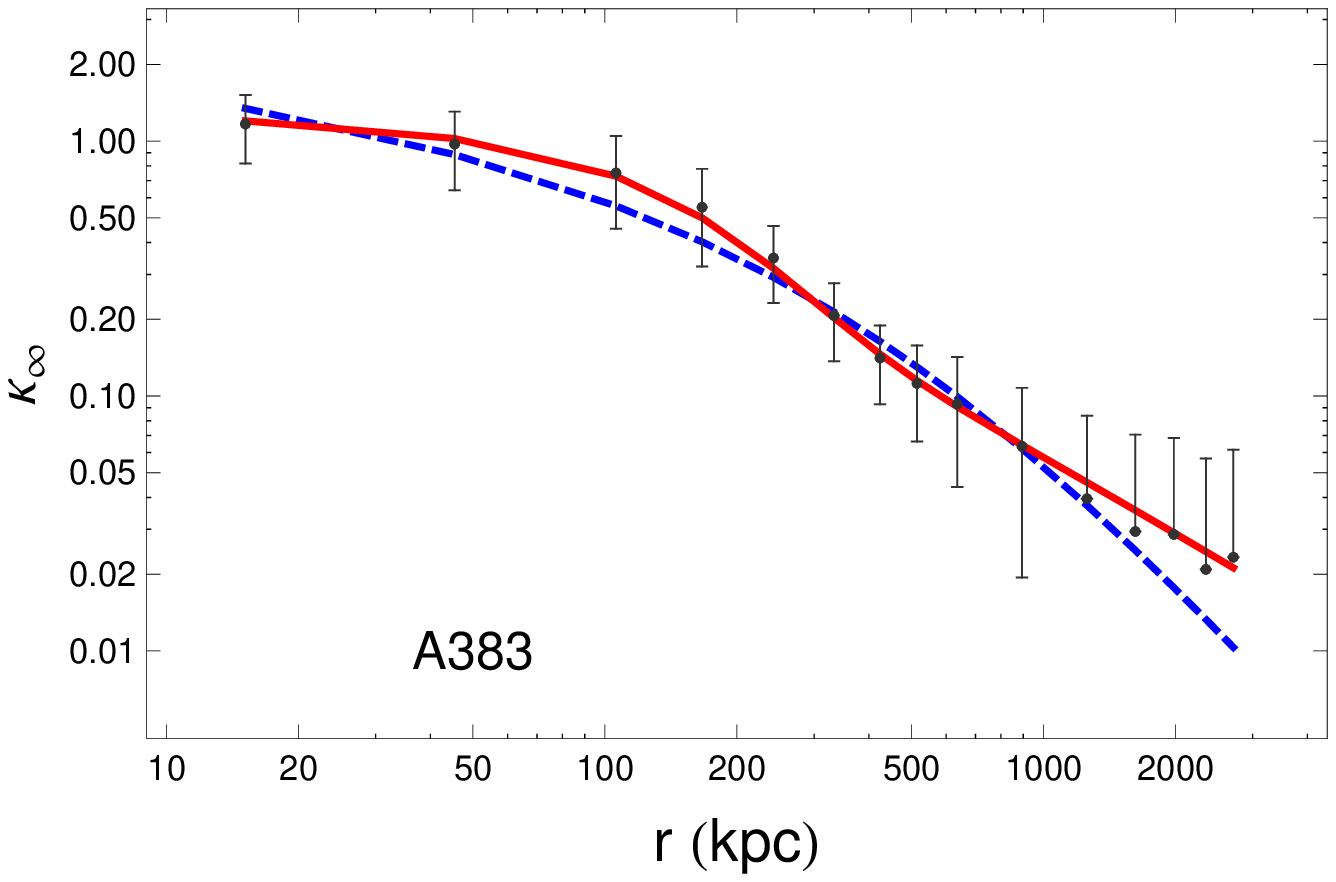}~
\includegraphics[width=7.6cm]{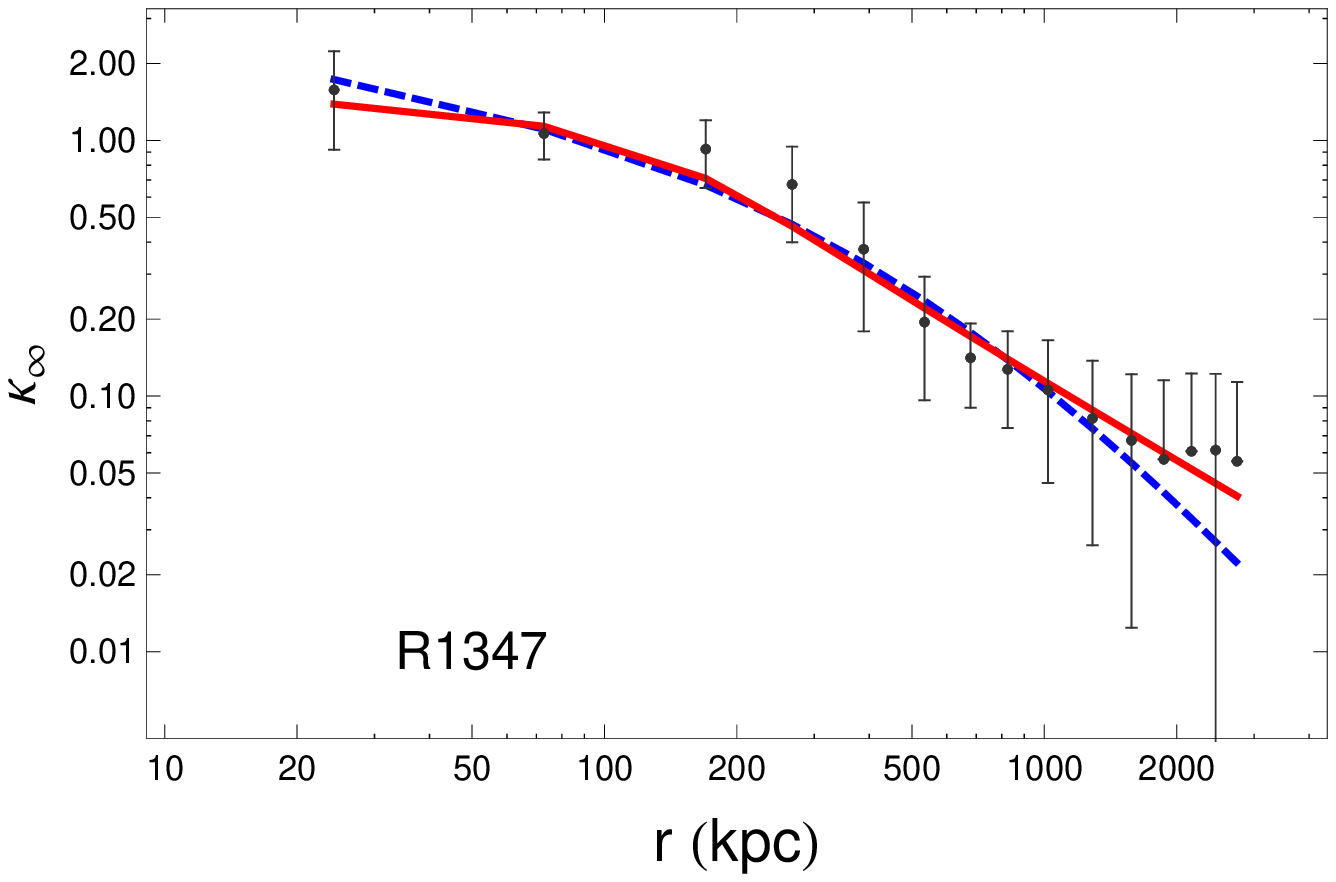}\\
~~~\\
\includegraphics[width=7.6cm]{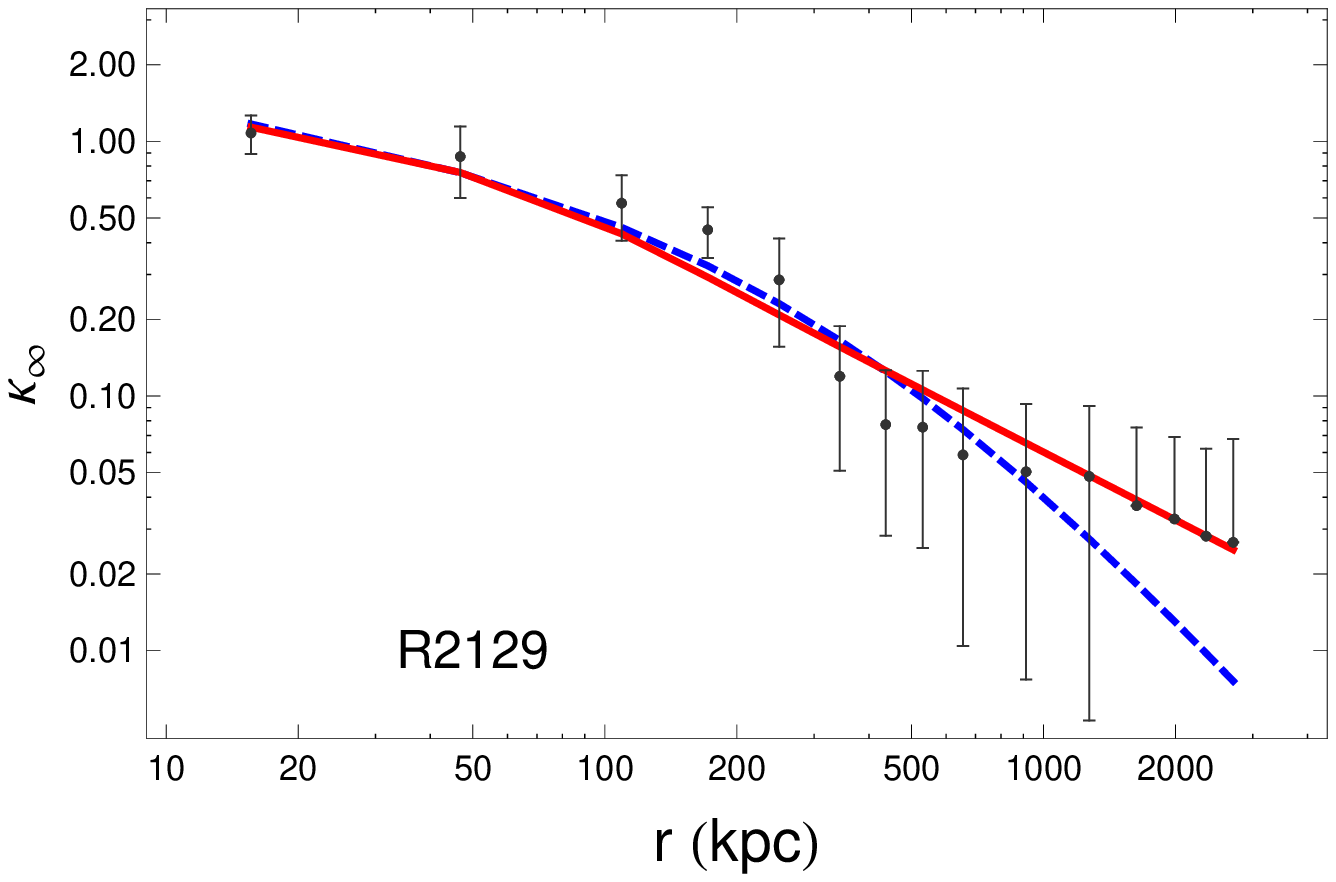}~
\includegraphics[width=7.6cm]{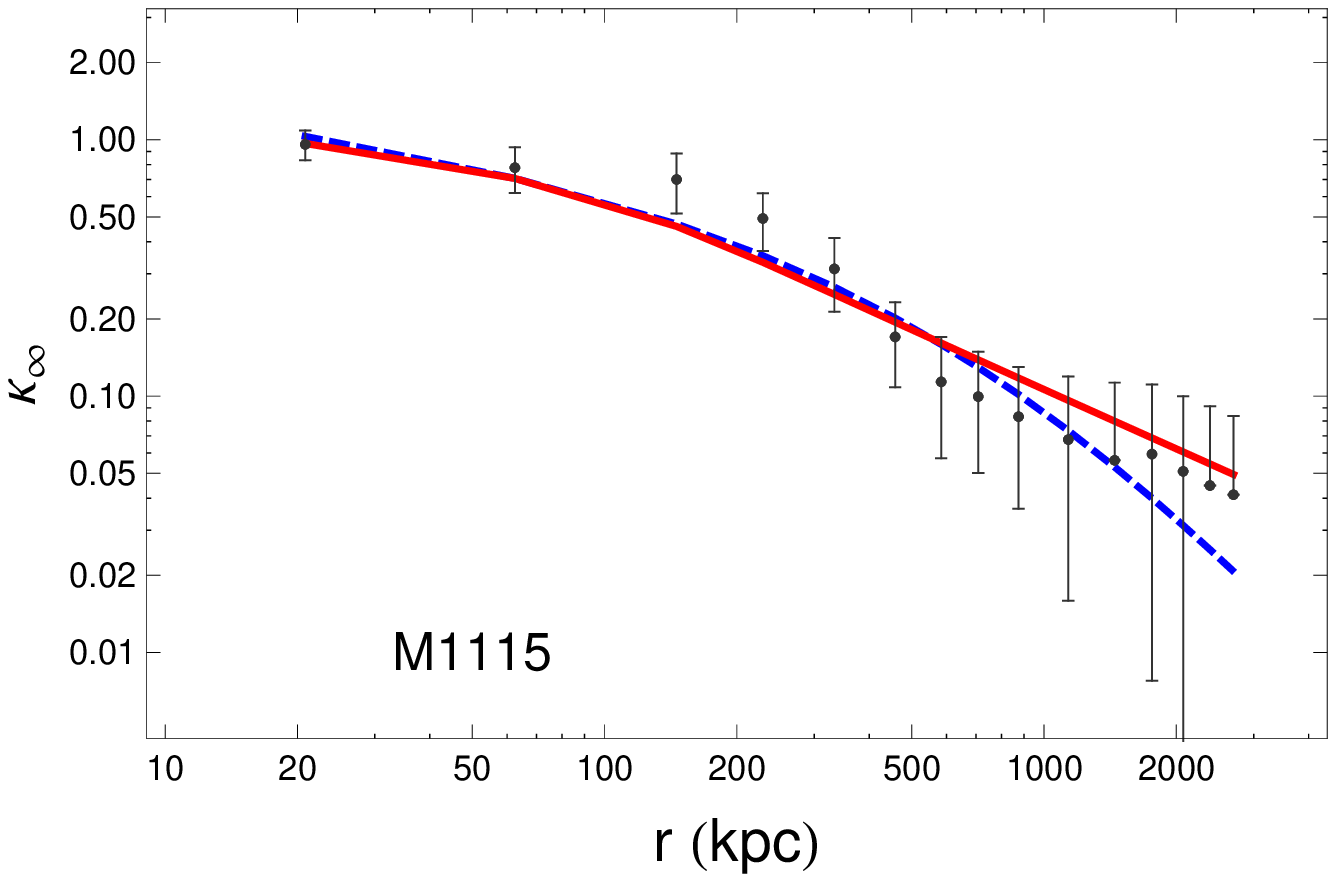}
\caption{Convergence map from gravitational lensing reconstruction. Color code: grey points - observational data; dashed blue - GR + NFW; solid red - Galileon + gas.}\label{fig:fit_2}
\end{figure*}

\begin{figure*}[htbp]
\centering
\includegraphics[width=7.6cm]{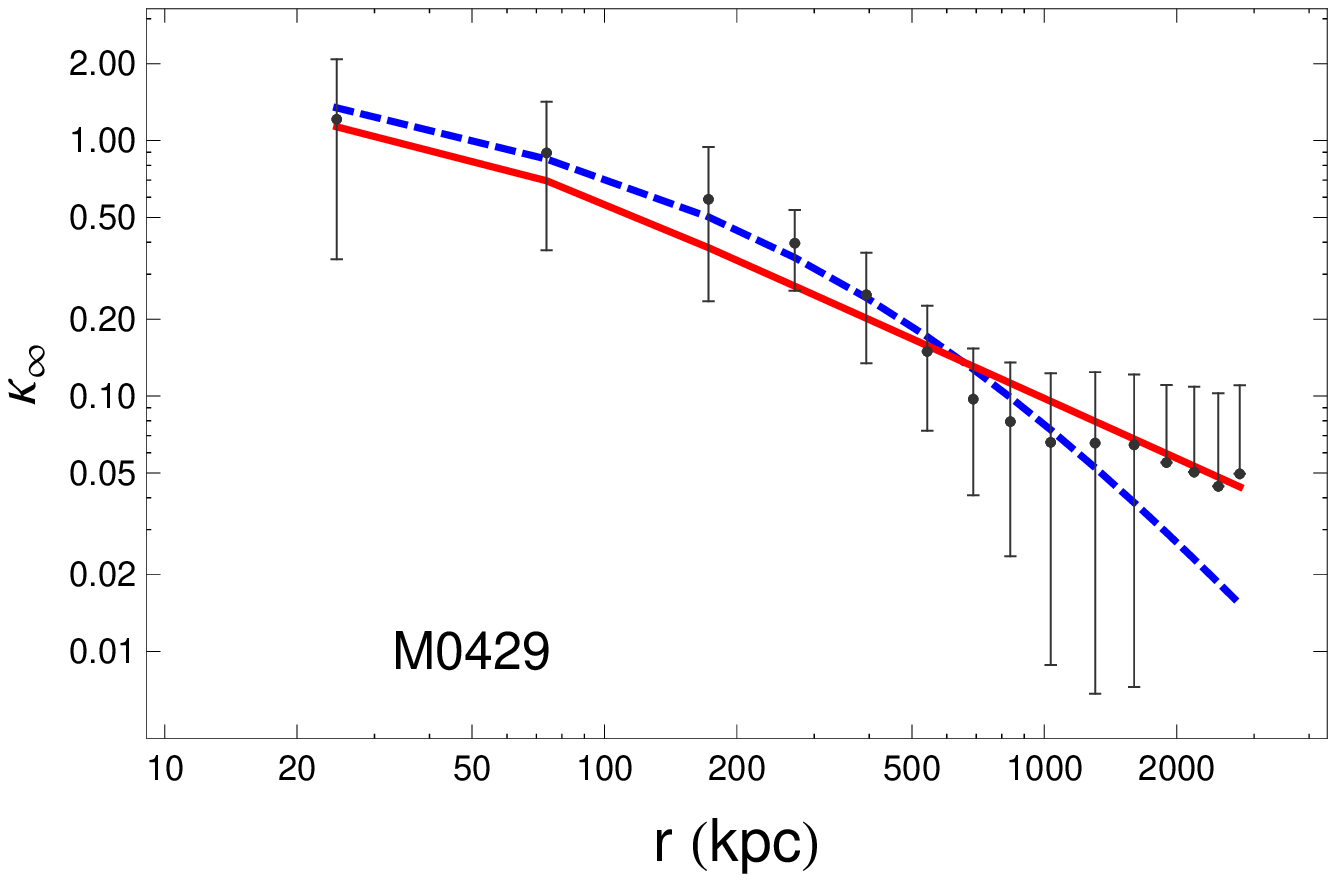}~
\includegraphics[width=7.6cm]{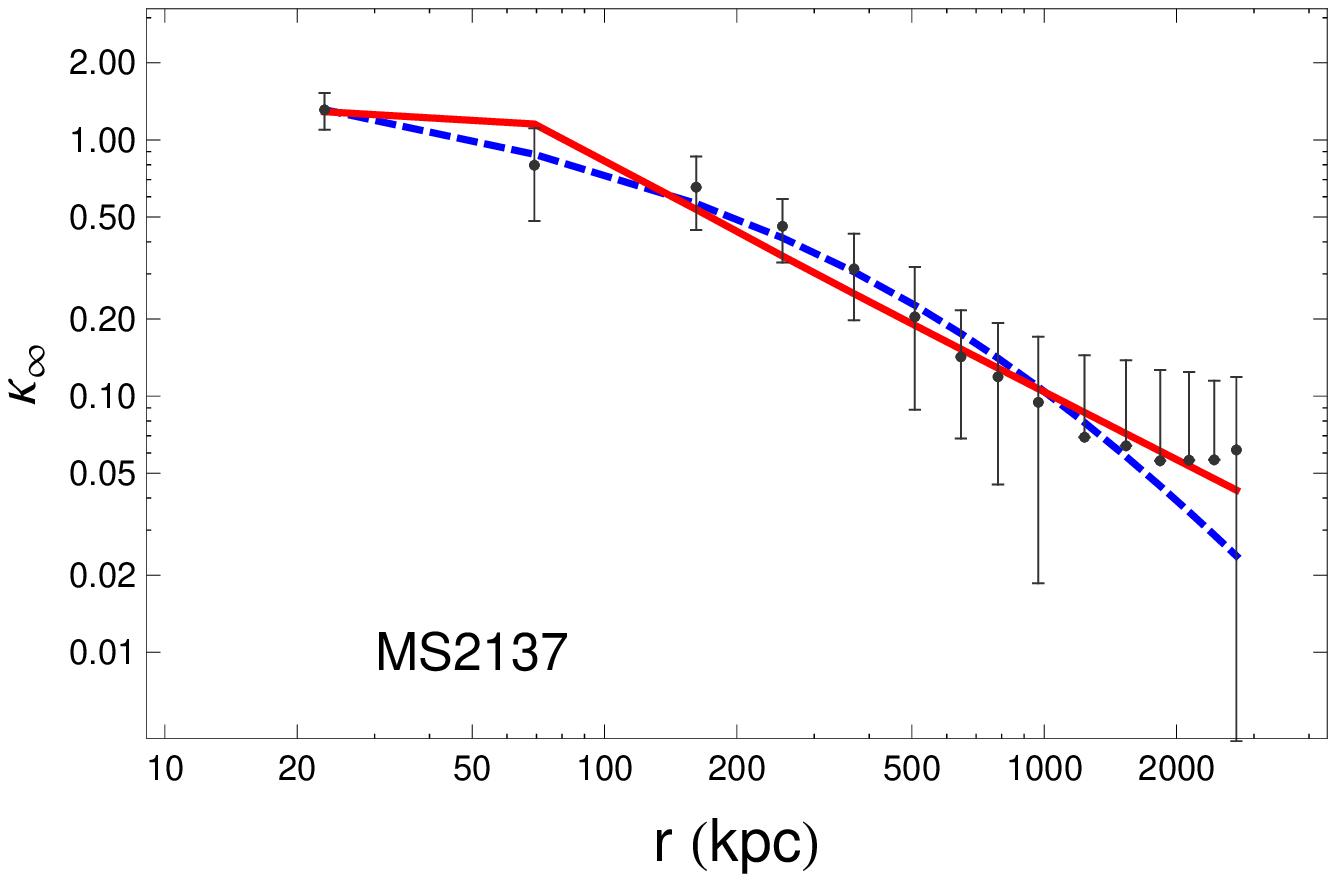}\\
~~~\\
\includegraphics[width=7.6cm]{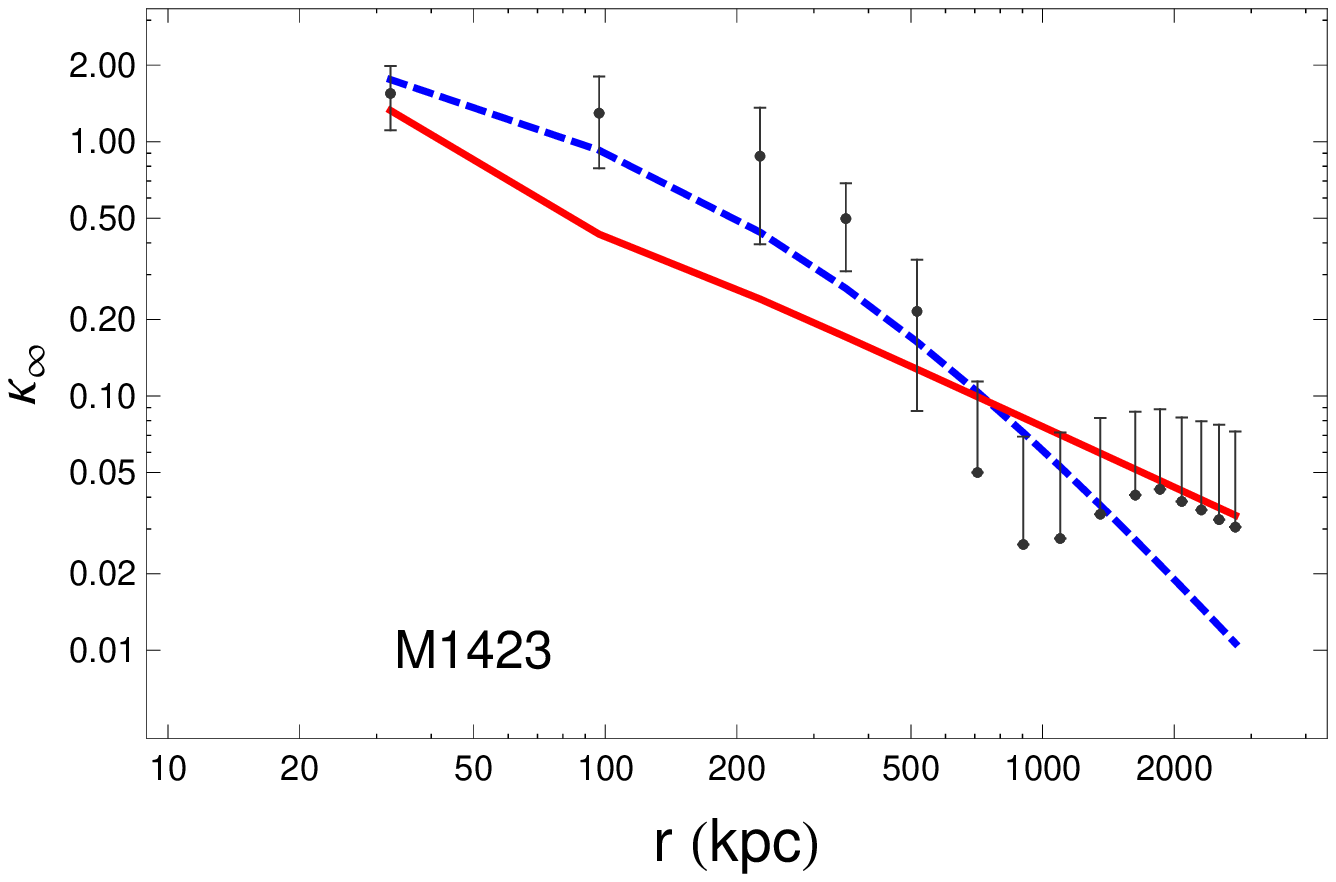}~
\includegraphics[width=7.6cm]{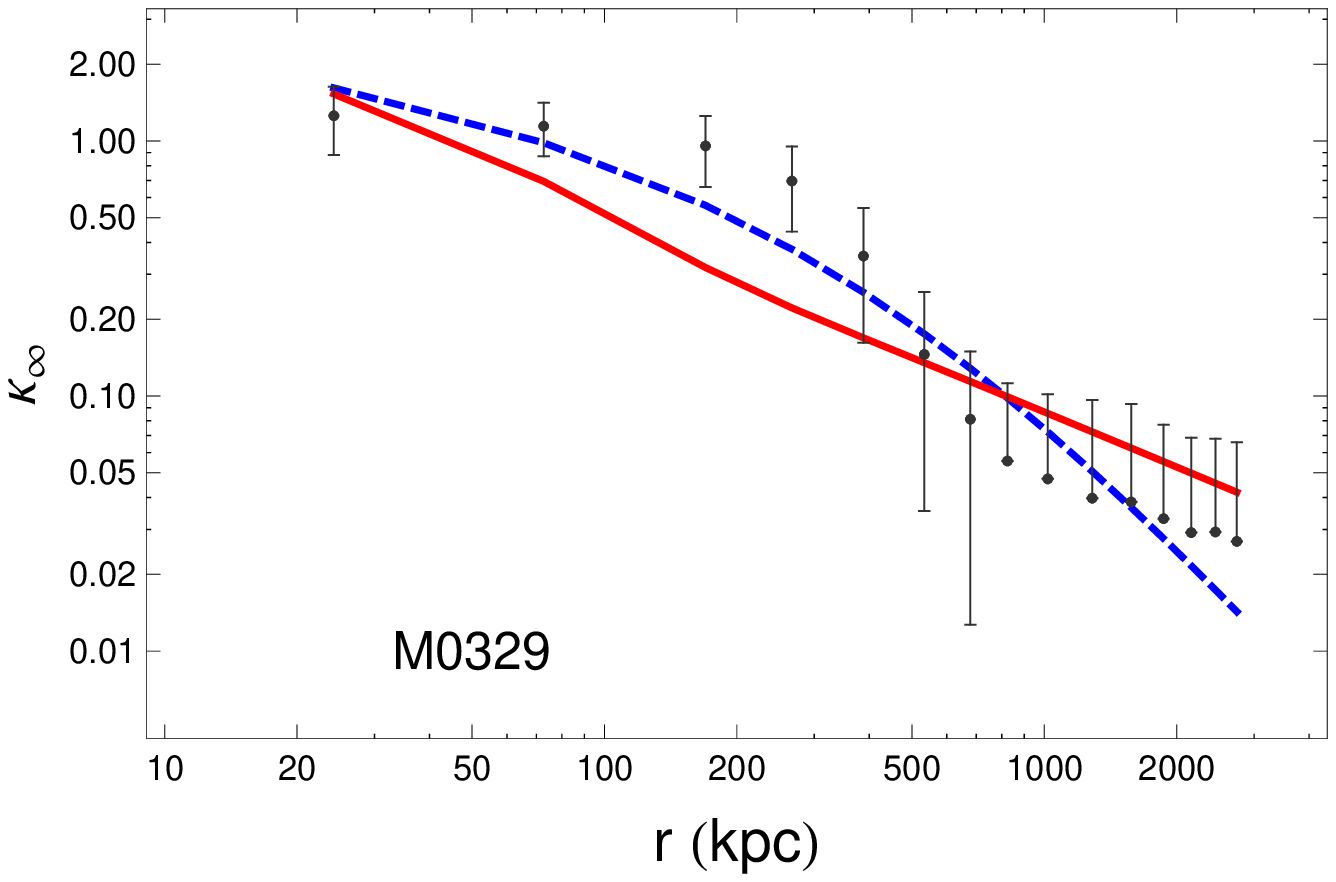}
\caption{Convergence map from gravitational lensing reconstruction. Color code: grey points - observational data; dashed blue - GR + NFW; solid red - Galileon + gas.}\label{fig:fit_3}
\end{figure*}

All our results are shown in Table~\ref{tab:results} and in Figs.~\ref{fig:fit_1}, \ref{fig:fit_2} and \ref{fig:fit_3}. First, we want to point out the results from the classical GR scenario are in perfect agreement with those shown in \citep{Merten15}; thus, our algorithm is performing well. Then, in Table~\ref{tab:results}, we have divided the clusters in three groups; this division was in some way done \textit{a posteriori}, in the effort to understand and interpret the results in a consistent way:
\begin{itemize}
  \item first group: clusters with high entropy which are, thus, more relaxed. In this case the gas density should be less perturbed by astrophysical processes and match in a quite good way the gravitational potential from the cluster;
  \item second group: clusters with much lower entropy (thus farer from relaxation) and less than the $\lesssim 30$ keV cm$^{2}$ threshold which could be associated to the presence of strong cooling cores;
  \item third group: for these clusters we have too few points covering a too small range in distance in order to obtain a reliable fit of the gas profiles.
\end{itemize}
Each group is ordered by decreasing Bayesian ratio. In the last column of Table~\ref{tab:results} we report the range covered by the X-ray data; this is an important point because while lensing observations generally span a very extended interval, approximately from $20$ kpc to $2$ Mpc, the X-ray observations, instead, can be much more limited, ranging from $10$ kpc to maximum $900$ kpc (with differences among clusters in the sample). In particular, one should consider that the greatest perturbations to the dynamical relaxed status of the gas can come from the very inner regions; and this may have some influence on the results. This is actually what we have found out: for all clusters from the first group which, as said, are the most relaxed, we have been able to perform the fit of the gas density and use gas data from all the available ranges. This is not the case for clusters with lower values of entropy, and thus farer from the relaxed state: for clusters from group two, for example, the Galileon model fails to match lensing data if the fit for the gas density is performed over all the given range. Instead, if we cut the very inner regions (generally, distances $<50$ kpc), the Galileon model matches the convergence data in a very performing way, as good as GR or even better. Clusters from group three, unfortunately, have too few points and distributed over a much limited distance range from clusters' centers, for this inner-regions-cut-procedure to be applied and to obtain some reliable result.

It is not strange to expect such behavior from the Galileon and not from GR, at least in the case we are considering here, i.e. where the only contribution to the cluster mass comes from gas. In GR, we are considering a dominant component ($6-8$ times more than gas), the NFW DM profile, which is quite insensitive to internal dynamics, so that it has no problem to fit data also in the most internal regions. On the other hand, in the Galileon case we are using directly gas mass, which might be influenced by it.

If we give a much detailed look to the $\chi^2$ and Bayesian Factor values, one can easily find out that we are able to fit the convergence maps from lensing considering only gas contribution, and with no need of dark matter at all. In most of the cases the Bayesian Factor is $<|1|$, which means that, even if there is no strong evidence in favour of one model than another, still the Galileon approach is as statistically valid as GR. And with one additional property whose weight cannot be properly taken into account by any statistical tool: within the Galileon model we are using only observed (baryonic) matter components plus some new scalar field (whose existence has to be checked, anyhow) which might play both the role of dark matter at clusters' scales, and of dark energy at cosmological ones, while in GR we have to assume, at least, the presence of two different dark components in order to explain observations.

Furthermore, one can see how there are some clusters with a Bayesian Factor larger than one, which means that there is a substantial evidence in favour of the Galileon model with respect to GR. From Figs.~(\ref{fig:fit_1}), (\ref{fig:fit_2}) and (\ref{fig:fit_3}) one can see where the Galileon behaves better than GR: it fits better low-medium distances $\approx 100-200$ kpc (it seems to better follow the decreasing trend of the convergence), and also very large distances $>1$ Mpc. Anyway, the latter have less statistical weight because of the larger errors; the main improvement comes from the low-medium distance range.

\begin{figure*}[htbp]
\centering
\includegraphics[width=\textwidth]{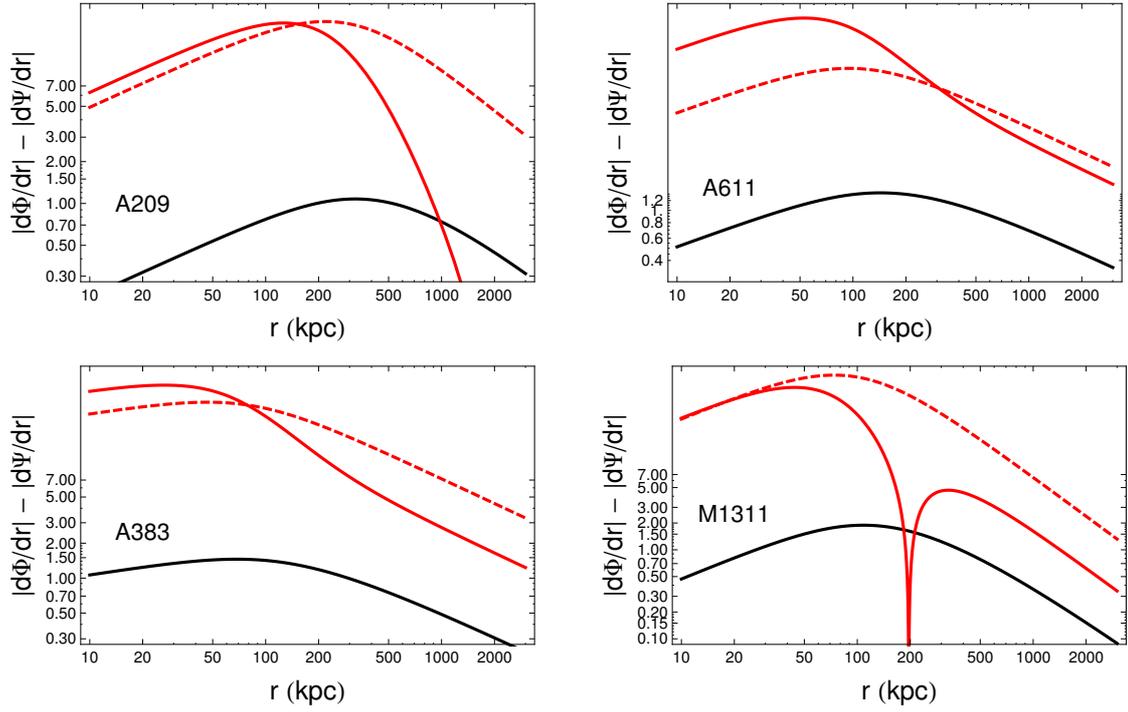}
\caption{Comparison of the classical Newtonian term ($\Upsilon_1$ and $\Upsilon_2$ tending to zero) which appears in both Eq.~\ref{eq:potential_phi} and Eq.~\ref{eq:potential_psi} (solid black line), with the characteristic Galileon terms, derived from the breaking of the Vainshtein screening mechanism, i.e. the terms proportional to $\Upsilon_1$ and $\Upsilon_2$ in, respectively, Eq.~\ref{eq:potential_phi} (solid red line) and Eq.~\ref{eq:potential_psi} (dashed red line).)}\label{fig:breaking_scale}
\end{figure*}

If we now center on the values for the Galileon parameters $\Upsilon_{1}$ and $\Upsilon_{2}$, we first have to note down that while in theory the parameters $\Upsilon_1$ and $\Upsilon_2$ should be the same for every cluster, actually, we expect some differences, due to the fact that we are comparing clusters at different evolutionary stages, which might translate in internal intrinsic dynamical differences affecting the mass densities and, thus, the measured values of these parameters. Then, we can verify the the values we find out are completely different from those in \citep{Sakstein2016}. Anyway, this is quite expected, as we have not used the stacked-profile approach which, in our opinion, is not the most suitable choice for the analysis of such kind of models which are characterized by new terms (like those in Eqs.~(\ref{eq:potential_phi}) and (\ref{eq:potential_psi})) which may depend in a stronger way than GR from possible intrinsic and peculiar properties, different for each cluster. Moreover, it should be noted that while the values from \citep{Sakstein2016} are still consistent with GR, namely, they are consistent with zero, in our case, instead, the GR limit is not strictly properly defined. Even if we had $\Upsilon_{1}\rightarrow 0$ and $\Upsilon_{2}\rightarrow 0$, we would have an ``incomplete'' GR unable to fit data, because lacking the main ingredient which makes it successful, DM. When discussing our results, there is another important caveat to bring in mind: when the galileon is compared with GR, the scale of the screening mechanism (or of its breaking) could be found out easily by checking where the new additional terms depending on $\Upsilon_1$ and $\Upsilon_2$ become important with respect to the ``classical'' Newtonian ones. But this would have sense only when comparing ``GR+dark matter+baryons'' with ``galileon+dark matter+baryon''. Here, we are exploring a completely different possibility, where the correction terms can play the role of dark matter over the entire astrophysical scale range we have been considering. Actually, the $\Upsilon_1$ and $\Upsilon_2$ terms in the potentials are important at all scales, otherwise they would not be able to replace dark matter wherever inside the cluster. As a further proof for such qualitative consideration, we have produced Fig.~\ref{fig:breaking_scale}. There, we compare the classical Newtonian term (e.g., as if both $\Upsilon_1$ and $\Upsilon_2$ were zero) which appear in both Eq.~\ref{eq:potential_phi} and Eq.~\ref{eq:potential_psi}, assuming only-gas mass profile (solid black line) with the characteristic Galileon terms, due to the breaking of the Vainshtein screening mechanism, i.e. the terms proportional to $\Upsilon_1$ and $\Upsilon_2$ in, respectively, Eq.~\ref{eq:potential_phi} (solid red line) and Eq.~\ref{eq:potential_psi} (dashed red line). It is clear that the Galileon terms are always (i.e. over the full range covered by data) dominant with respect to the classical term.

Thus, there would be a natural concern about whether such results are in contrast with Solar System constraints. In a naive way, Fig.~\ref{fig:breaking_scale} is comparing the classical Newtonian force exerted by only-visible matter with the fifth force generated by the Galileon field. In a classical Galileon theory, one would expect a scale separating the range where this ratio is $>1$ from that one where it is $<<1$, where the screening is effective; such scale should be cosmological. In Beyond-Horndeski theories with a breaking of the screening, one would still expect such scale, but now at smaller (maybe) astrophysical scales, so that, in principle, one could have observational signatures. In our case, instead, we apparently have no such scale, thus meaning that, in principle, the screening is broken all over the scale we are testing. While this result is somewhat expected, because we want to mimic dark matter with such a new mechanism, and dark matter extends all over the full extent of clusters of galaxies, on the other hand we have to find a way to reconcile it with local gravity experiments, which poses strong constraints on possible deviation from Newtonian force.

What we have found is that it is in principle possible to explain dark matter as a result of Vainshtein and GR breaking on scales as large as those tested ($100$ kpc $- 2$ Mpc). On smaller scales, in order to preserve GR, we need to fill the gap, i.e. to analyze smaller structures and find, lately, a gravitational structure which does not require $\Upsilon$ large but, eventually $\sim 0.1$ (this value is, for example, in \citep{KoyamaSakstein2015,SaksteinPRL15,Sakstein2016}, where the Galileon is considered - only - as a cosmological scale fluid) or less to be explained. Finally, we are just saying that the breaking scale might not be at cosmological scales (with Galileon playing the role of dark energy only), but at smaller ones too, albeit, much larger than Solar System.

We have also to point out that we are not using central galaxies contribution to the mass profiles of our clusters, and galaxies might play an important role in describing the cluster potentials in the very inner regions. Their can potentially lower the values of the constants $\Upsilon_1$ and $\Upsilon_2$ and, thus, reduce the contributions from the Galileon in such regions. Moreover, they will surely change the trend of the fifth force toward those inner regions, possibly making the fifth force drop even faster than present results.

\begin{figure*}[htbp]
\centering
\includegraphics[width=8.cm]{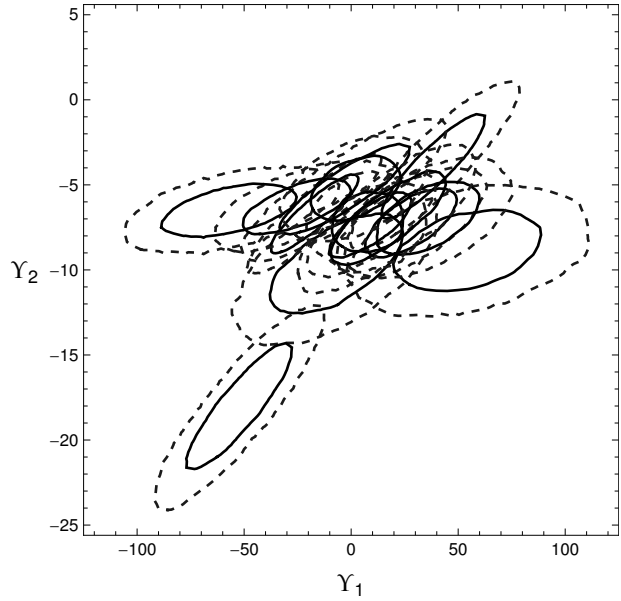}
\caption{Confidence contours for the Galileon parameters $\Upsilon_{1}$ and $\Upsilon_{2}$ from all clusters in the sample (except last four in Table~\ref{tab:results}).)}\label{fig:contour}
\end{figure*}

Given this difference, it is interesting to note from Table~\ref{tab:results} and from Fig.~(\ref{fig:contour}) that $\Upsilon_{2}$ is not consistent with zero, at least at $2\sigma$ confidence level, for all the clusters in our sample. The other parameter $\Upsilon_{1}$ spans a much wider range of possible values. We can also note how the relative errors on these parameters are much smaller than those reported in \citep{Sakstein2016}. Finally, if some cluster-features dependence is on-going, no clear trend is arguable with present data.

\section{Discussion and Conclusions}
\label{sec:Conclusions}

In this work we have tried to answer the question: is it possible to explain clusters of galaxies lensing observations without resorting to DM? This is clearly impossible in the classical context of GR where, actually, DM is vital and needed in order to explain observations. But it might be possible in some alternative models of gravity. Most of those have been introduced to explain large scale effects of DE; then, they need some screening mechanism to turn off the new forces/degrees of freedom at such scales where GR should be recovered. The same screening mechanisms are the main obstacle to elect many alternative models as DM sources too, because the suppression of the new degrees of freedom generally happens at astrophysical scales. A safety alternative come from models like that described in \citep{KoyamaSakstein2015,Sakstein2016}, a modified Galileon-type model where the screening mechanism can be broken; in such a case, distinctive signatures of the new theory should be observable. And, we add, they might play the role of DM.

We have used the above model and applied it to a sample of $18$ clusters of galaxies observed by the spatial survey \textit{CLASH}. We have joined two different observational probes: mass reconstruction from lensing events (both strong and weak lensing); and X-ray observations of the hot intra-cluster gas. We have used the latter to derive the gas mass profile of such clusters; and we have used this information trying to reproduce the former observations. The key requirement in our analysis is that the mass of the clusters is made only by gas, no DM is considered at all. Anyway, one should not forget that baryonic matter in clusters is also made of galaxies. Unfortunately, we have no data available for the sample we considered in this sense. Including galaxies, of course, might affect our results: in particular, galaxies could play a larger role in inner regions, with a consequent ``re-normalization'' of the mass profile, and possible influence on the theoretical parameters.

Results show that when the screening mechanism is broken, then the Galileon model can be used to match DE at large cosmological scales, and DM at smaller ones. In particular, it results that the Galileon model is even more statistically favourable than the GR to match lensing observations. Far from us to state that this ends the DM question, or that such model is ``the'' model which can win GR. But we think this analysis is useful to state that much more attention should be paid to GR alternative and competing models \cite{rev}, because they can be as much successful as GR. And it could help theoreticians to understand what is the right path to follow in order to build/discover the true underlying gravity theory behind our universe.

\section*{Acknowledgments}

The research of V.S. and M.P.D. was financed by the Polish National Science Center Grant DEC-2012/06/A/ST2/00395. D.F.M. thanks the Research Council of Norway and the NOTUR computer facilities. S.C. acknowledges financial support of INFN (iniziative specifiche TEONGRAV and QGSKY). This article is also based upon work from COST action CA15117 (CANTATA), supported by COST (European Cooperation in Science and Technology).

{\renewcommand{\tabcolsep}{1.mm}
{\renewcommand{\arraystretch}{1.75}
\begin{table*}
\begin{minipage}{\textwidth}
\centering
\caption{Primary NFW parameters from separate fits for X-ray and lensing data. Units: densities are in $10^{15} \, M_{\odot}$ Mpc$^{-3}$; masses are in $10^{14}\, M_{\odot}$; radii are in Mpc. For the gas range, the $x\rightarrow y$ means that we have changed the minimum distance from the center for the gas density fit from the original value $x$ to $y$, in order to avoid influences from non-gravitational phenomena.}\label{tab:results}
\resizebox*{\textwidth}{!}{
\begin{tabular}{c|ccc|ccc|cccc}
  \hline
  \hline
  name  & \multicolumn{3}{c|}{GR + NFW}          & \multicolumn{3}{c|}{Galileon + Gas}                  & \multicolumn{2}{c}{Evidence} & Entropy\footnote{Core entropy in keV cm$^{2}$; if $\lesssim 30$ then there is strong cooling core. Data are from \citep{Donahue14}.} & Gas range \\
        & $\rho_{s}$ & $r_{s}$ & $\chi^{2}_{GR}$ & $\Upsilon_{1}$ & $\Upsilon_{2}$ & $\chi^{2}_{Gal.}$ & $\mathcal{B}^{Gal.}_{GR}$ & $\ln \mathcal{B}^{Gal.}_{GR}$ & $K_{0}$ & $[R_{min}, R_{max}]$ \\
  \hline
  \hline
  A209 & $0.639^{+0.417}_{-0.257}$ & $0.601^{+0.212}_{-0.152}$ & $2.87$ & $28.88^{+13.88}_{-13.82}$ & $-7.11^{+1.30}_{-1.29}$ & $0.48$ & $3.35$ & $1.21$ & $105.5$ & $[0.093, 0.528]$ \\
  MACSJ1206 & $1.561^{+1.301}_{-0.694}$ & $0.401^{+0.141}_{-0.113}$ & $4.86$ & $8.82^{+10.65}_{-11.27}$ & $-7.13^{+1.15}_{-1.11}$ & $3.25$ & $2.31$ & $0.84$ & $69.0$ & $[0.037, 0.847]$ \\
  MACSJ1720 & $1.480^{+1.219}_{-0.682}$ & $0.389^{+0.151}_{-0.112}$ & $4.22$ & $36.61^{+15.57}_{-15.52}$ & $-6.84^{+1.38}_{-1.38}$ & $2.56$ & $2.31$ & $0.84$ & $94.4$ & $[0.023,0.422]$ \\
  A2261 & $0.649^{+0.565}_{-0.312}$ & $0.634^{+0.292}_{-0.192}$ & $3.69$ & $-5.63^{+19.49}_{-19.19}$ & $-9.52^{+1.84}_{-1.86}$ & $2.17$ & $2.06$ & $0.72$ & $61.1$ & $[0.023\rightarrow0.056,0.708]$ \\
  MACSJ1311 & $1.662^{+0.786}_{-0.511}$ & $0.315^{+0.081}_{-0.071}$ & $4.03$ & $-51.15^{+15.88}_{-15.57}$ & $-17.87^{+2.38}_{-2.39}$ & $2.95$ & $1.71$ & $0.54$ & $47.4$ & $[0.028, 0.257]$ \\
  A611 & $0.752^{+0.426}_{-0.269}$ & $0.541^{+0.180}_{-0.138}$ & $4.08$ & $39.81^{+15.67}_{-15.88}$ & $-3.72^{+1.98}_{-2.01}$ & $3.16$ & $1.62$ & $0.48$ & $125.0$ & $[0.023,0.800]$ \\
  RXCJ2248 & $0.923^{+1.029}_{-0.530}$ & $0.547^{+0.315}_{-0.186}$ & $1.31$ & $2.95^{+13.30}_{-13.55}$ & $-5.14^{+1.22}_{-1.21}$ & $0.70$ & $1.37$ & $0.32$ & $42.2$ & $[0.029,0.751]$ \\
  MACSJ0744 & $1.921^{+1.362}_{-0.731}$ & $0.346^{+0.111}_{-0.093}$ & $3.16$ & $10.36^{+11.52}_{-12.46}$ & $-7.25^{+1.17}_{-1.18}$ & $3.23$ & $1.03$ & $0.03$ & $42.4$ & $[0.046,0.679]$ \\
  \hline
  \hline
  MACSJ1931 & $1.242^{+1.907}_{-0.743}$ & $0.359^{+0.260}_{-0.152}$ & $4.18$ & $-1.53^{19.23}_{-19.30}$ & $-5.15^{+1.69}_{-1.68}$ & $1.33$ & $4.65$ & $1.54$ & $14.6$ & $[0.010\rightarrow 0.045,0.940]$\\
  RXJ1532 & $0.692^{+0.517}_{-0.315}$ & $0.485^{+0.198}_{-0.136}$ & $6.90$ & $-55.19^{+19.79}_{-19.97}$ & $-6.40^{+0.98}_{-0.98}$ & $4.78$ & $2.91$ & $1.07$ & $16.9$ & $[0.008\rightarrow 0.044,0.732]$ \\
  A383 & $1.405^{+0.986}_{-0.638}$ & $0.434^{+0.162}_{-0.108}$ & $2.05$ & $56.16^{+22.04}_{-21.96}$ & $-8.75^{+1.53}_{-1.51}$ & $0.23$ & $2.50$ & $0.92$ & $13.0$ & $[0.012,0.451]$ \\
  RXJ1347 & $0.867^{+0.621}_{-0.381}$ & $0.572^{+0.204}_{-0.148}$ & $3.23$ & $-23.74^{+16.06}_{-16.51}$ & $-6.14^{+1.04}_{-1.05}$ & $2.16$ & $1.69$0 & $0.53$ & $12.5$ & $[0.011\rightarrow 0.052,0.685]$ \\
  RXJ2129 & $1.227^{+0.747}_{-0.463}$ & $0.386^{+0.125}_{-0.098}$ & $5.30$ & $17.92^{+17.80}_{-16.79}$ & $-6.87^{+1.72}_{-1.71}$ & $5.46$ & $0.92$ & $-0.08$ & $21.1$ & $[0.012\rightarrow 0.072,0.864]$ \\
  MACSJ1115 & $0.341^{+0.156}_{-0.101}$ & $0.818^{+0.197}_{-0.174}$ & $5.62$ & $16.30^{+12.04}_{-12.26}$ & $-6.59^{+1.35}_{-1.35}$ & $6.36$ & $0.71$ & $-0.34$ & $14.8$ & $[0.016\rightarrow 0.070,0.324]$ \\
  \hline
  \hline
  MACSJ0429 & $1.277^{+2.251}_{-0.859}$ & $0.424^{+0.307}_{-0.170}$ & $1.87$ & $-18.76^{+88.78}_{-91.84}$ & $-7.41^{+4.68}_{-4.80}$ & $2.51$ & $0.77$ & $-0.26$ & $17.2$ & $[0.018,0.324]$ \\
  MS2137 & $0.564^{+0.349}_{-0.223}$ & $0.728^{+0.256}_{-0.189}$ & $1.52$ & $-110.09^{+39.89}_{-42.33}$ & $-18.94^{+3.42}_{-3.54}$ & $2.86$ & $0.52$ & $-0.66$ & $14.7$ & $[0.012,0.199]$ \\
  MACSJ1423 & $2.472^{+2.363}_{-1.248}$ & $0.283^{+0.125}_{-0.091}$ & $6.44$ & $235.69^{+551.34}_{-545.35}$ & $3.52^{+23.70}_{-23.56}$ & $11.75$ & $0.07$ & $-2.65$ & $10.2$ & $[0.008,0.160]$ \\
  MACSJ0329 & $1.246^{+0.919}_{-0.536}$ & $0.416^{+0.153}_{-0.116}$ & $6.35$ & $62.50^{+46.06}_{-45.64}$ & $-0.88^{+3.34}_{-3.34}$ & $14.88$ & $0.01$ & $-4.2$ & $11.1$ & $[0.023,0.195]$ \\
\hline
\hline
\end{tabular}}
\end{minipage}
\end{table*}}}

\clearpage
\newpage

\end{document}